# Have wind and solar energy increased electricity price volatility?

*And what is electricity price volatility, anyway?*

Lion Hirth

Hertie School, Berlin, and Neon Neue Energieökonomik, Berlin

hirth@hertie-school.org

**Abstract:** It is widely held that wind and solar power increase the volatility of electricity prices. This paper argues that the literature conflates two distinct types of price variation: across delivery time and across trading time. Determined simultaneously in the same day-ahead auction, prices for power to be delivered in distinct quarter-hours differ. This price dispersion, which I call "variability", is rooted in non-storability. After that auction, trading continues in intraday markets until imbalance settlement. The price for the same delivery period changes as new information arises, e.g., weather forecasts. Such price "volatility" is ultimately caused by uncertainty. This paper develops a simple algebraic model of day-ahead and imbalance markets that provides a formal definition of variability and volatility, allows for a decomposition and derives the fundamental drivers, uncertainty and non-storability. It then provides empirical data on price variability and volatility for the past 25 years in Germany, during which the share of wind and solar energy increased from zero to nearly 50 % of generation. Volatility is measured with imbalance prices, intraday price indices and intraday transaction data. Variability followed a U-shaped trajectory, falling from the 2000s to the 2010s but increasing since then. Volatility, in contrast, declined over time. Across 24 European bidding zones, variability has risen since the late 2010s in almost all zones and increases with the share of solar power.

# 1. Introduction

It is a widely shared view among market participants, policy makers and researchers that wind and solar power make electricity prices more volatile. The intuition is simple: prices are depressed when it is sunny or windy and jump to high levels when it is not. Empirical studies of renewables and price "volatility" frequently find such effects (Ketterer 2014; Johnson and Oliver 2019; Cevik and Ninomiya 2023), although not uniformly (Rintamäki et al. 2017; Schöniger and Morawetz 2022).

This paper argues that the narrative conflates two phenomena. In finance, volatility describes how the price of a given asset changes over time as new information arrives. Electricity day-ahead prices, in contrast, are mostly discussed as a sequence of prices for different delivery periods, each of which is a different good because electricity cannot be stored economically. Price differences between delivery periods arise even if all information is known in advance. I call these differences variability and reserve the term volatility for price changes of the same delivery period across trading time. Writing $p(\tau, t)$ for the price at trading time $\tau$ for delivery period $t$, volatility is dispersion along $\tau$, variability dispersion along $t$.

How have wind and solar power changed variability and volatility? How are the two related? Answering these questions requires a clear separation of the two concepts, a measurement strategy for each and data that cover a long period of renewable expansion.

The paper makes four contributions.

- First, it introduces the distinction between "price variability" and "price volatility" as two distinct types of price dispersion in electricity markets. It identifies their necessary conditions and relates the terms to their use in the finance and integration cost literature.
- Second, it develops a simple algebraic model of day-ahead and imbalance markets that provides a formal definition of variability and volatility, allows for a decomposition and derives the fundamental drivers, uncertainty and non-storability.
- Third, it reviews the literature and maps the research to the two concepts.
- Fourth, it provides empirical data on price variability and volatility for the past 25 years in Germany, during which the share of wind and solar energy increased from zero to nearly 50 % of generation. To measure volatility, it uses imbalance prices, intraday price indices and transaction data. Variability followed a U-shaped trajectory, falling from the 2000s to the 2010s but increasing in the 2020s. Volatility, in contrast, declined over time. For 24 European bidding zones, it compares variability since 2015 and relates it to the shares of wind and solar power.

In Germany, normalized variability fell from the 2000s to the 2010s and has risen since 2020; since 2024, it has been at the level of the 2000s. Normalized volatility has not increased; transaction data for 2015–2025 confirm this. Volatility against the imbalance price has declined since 2015. Since 2024, variability has been highest on days with high solar generation and lowest on windy days; volatility no longer rises with wind generation. In absolute terms, both variability and volatility rose with the price level after 2021. Across Europe, variability was higher in 2024–2026 than in 2015–2019 in 21 of 22 bidding zones. Monthly variability rises with the share of solar power.

The first part of the paper is conceptual: Section 2 introduces the concepts, Section 3 reviews the literature and Section 4 provides a formal model. The second part is empirical: Section 5 describes data and measures and Section 6 presents the results for Germany and compares variability across European bidding zones.

# 2. Electricity price volatility and variability

This section introduces what I call "price volatility" and "price variability" in short-term markets for electricity. It is mostly non-technical; a formal definition follows in Section 4.

## 2.1. Electricity: a different good every 15 minutes

Despite the advances in batteries, storing electricity remains much more expensive than storing other energy carriers. The cost of battery cells is on the order of USD 60 per kWh, compared to about USD 0.2 per kWh for natural gas storage and USD 0.03 per kWh for oil, i.e., several hundred or thousand times more. For the purpose of wholesale market modeling, it is often appropriate to consider electricity as non-storable.

As a consequence of non-storability, the market-clearing price is essentially determined by *instantaneous* demand and supply conditions, i.e., mostly independent of previous or future market conditions. This can be observed in the form of five price patterns mostly absent from other commodity markets:

- Day-ahead prices vary greatly over short time horizons, often hundreds of percent within the day. Adjacent hours can differ sharply, e.g., solar hours at midday and the evening peak. If electricity could be stored, arbitrage would prevent such wide swings.
- Day-ahead prices regularly become zero or turn negative for brief periods such as a few hours. If electricity could be stored, producers would not sell at negative prices. At the other end, prices spike to extreme levels in scarcity situations, sometimes several orders of magnitude above average prices. If electricity could be stored, inventories would buffer scarcity.
- Day-ahead prices fluctuate in predictable patterns even when no new information arrives, such as diurnal shapes that follow the regular movements of electricity consumption and solar energy. Daily, weekly and seasonal profiles are commonly known and recur every year. If electricity could be stored, predictable patterns would be exploited by traders and disappear.
- Viewed as a time series across delivery periods, day-ahead prices revert to their mean within hours or days instead of following a random walk. Prices of storable commodities also revert to a mean, but much more slowly (Lucia and Schwartz 2002; Weron 2006; Schwartz 1997). Mean reversion along delivery time thus describes variability, not the price path of one electricity good.
- Day-ahead and forward prices are not linked by a cost-of-carry relation. For storable commodities, storage arbitrage ties the forward price to the current price plus storage and financing costs. Electricity forward prices instead reflect expected day-ahead prices and a risk premium (Bessembinder and Lemmon 2002).

In European electricity markets, a different electricity price is determined for every 15-minute interval. This is the "market time unit", the native temporal granularity of electricity. Other markets use pricing intervals between 5 minutes and 1 hour.

Economically, electricity delivered in different MTUs is best understood as different economic goods. Electricity delivered in one quarter-hour is a different good from electricity delivered in the next quarter-hour, because one is a poor substitute for the other. They may be called individual "electricity goods" or "electricity products". The low degree of substitutability is the direct result of non-storability. Different goods have different prices. The fact that electricity delivered in each of the 35,040 quarter-hours of the year has a different price gives rise to what I call "price variability".

## 2.2. From day-ahead auctions to imbalance prices

Well-developed short-term markets for electricity have a sequence of market stages. In Europe, quarter-hourly electricity prices are first determined in the coupled European day-ahead auction, which takes place at noon the day before delivery and establishes the price for each of the 96 quarter-hours of the next day (Figure 1). Before that, electricity is traded in forward contracts, but those bundle multiple quarter-hours into strips and hence do not trade individual electricity goods (Section 2.4). Day-ahead auctions are double-sided sealed-bid auctions that apply marginal pricing and are organized by power exchanges such as Nord Pool or EPEX SPOT. They are multi-item auctions: 96 quarter-hourly contracts are auctioned at once (24 hourly contracts until September 2025; quarter-hourly contracts were introduced on 1 October 2025). Hence a one-year time series of day-ahead prices comprises 365 auction results determined one day apart, each of which holds 96 simultaneously determined quarter-hourly prices. In many ways, the day-ahead auction is the "lead market" for electricity: it has the largest volume of all short-term markets, its price is the underlying of financial forward contracts, and the auctions of most European countries are coupled through implicit market coupling such that most imports and exports are determined in that auction.

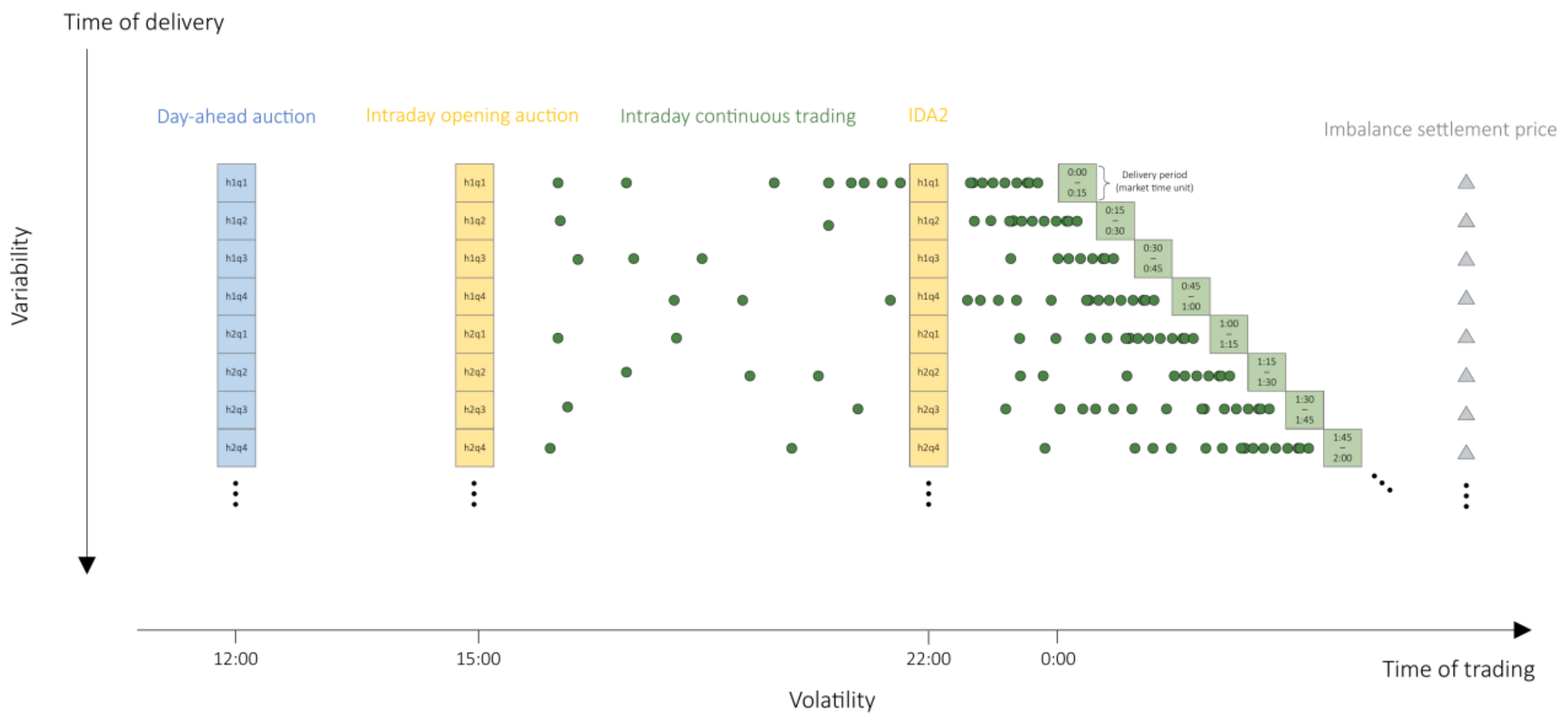

Figure 1. Trading time and delivery time of electricity products

The day-ahead auction, however, is not the end of the story; quarter-hourly electricity products continue to be traded. There is a series of three pan-European intraday auctions, taking place at 15:00 and 22:00 on the day before delivery and at 10:00 on the delivery day; the last one covers only the delivery periods from 12:00 to 24:00. In addition, continuous trading opens at 15:00 on the day before delivery and runs until shortly before the start of the delivery period. Gate closure for cross-border trades was 60 minutes before delivery until 2025; since 2026, EU rules require 30 minutes, with derogations possible until 2029. Within Germany, trading continues until 30 minutes before delivery and within a control area until 5 minutes before delivery. The last step is imbalance settlement. The imbalance settlement price is used to clear any remaining open positions of market parties, based on the physically metered volumes. This, too, happens at 15-minute granularity.

While some countries had a sequence of auctions, Germany traditionally used a continuous market to trade electricity until a few minutes before delivery. Quarter-hourly products were introduced in continuous trading in 2011. An auction just before the opening of the continuous market followed in 2014. Two additional auctions at later points in time were introduced in 2024. As a consequence of the continuous trading and multiple auctions, there is not a single intraday price for each quarter-hour. Continuous trading consists of many small transactions, on average about ten per second, and about two-thirds of trades take place in the last two hours before delivery.

When market parties end up being short or long of energy, they are financially settled at the imbalance settlement price, the real-time price of wholesale electricity. This price is determined by transmission system operators, based on complex formulas that may use day-ahead or intraday prices but also balancing energy prices as inputs, apply non-linear transformations and price floors and ceilings as well as multipliers and adders. System operators as well as national regulatory authorities have significant discretion to define these pricing rules, so that imbalance pricing rules are heterogeneous across countries. They may also include forms of price discrimination, differentiating between generators and loads, or between short and long positions. It is not a price that emerges from voluntary transactions among market parties, but an administratively determined one. In a way, however, it is the "ultimate" electricity price. All previous markets, including intraday and day-ahead prices, can be understood as forward markets, trading expectations of imbalance settlement prices. Settlement price formulas have changed regularly, such that they must be used with care in long-term trend analysis. Most short-term speculative trading is position-taking against the imbalance price: traders go long or short in the intraday market if they expect the imbalance price to exceed or fall short of the current intraday price.

Essentially, intraday markets and imbalance settlement are mechanisms to digest new information, such as updated weather forecasts (affecting renewable energy and electricity demand) and outages of interconnectors and generators. The change in electricity prices for the same delivery period between the first price (day-ahead auction) and later prices (intraday, imbalance price) is conventionally called price volatility.

I avoid the term "spot price" in this paper. In finance, the spot price is the price for immediate delivery, the end point of the trading timeline of a commodity and the underlying for cash-settled forward contracts. In European electricity markets, the term conventionally denotes the day-ahead price. However, the day-ahead price is not the last price, because it is followed by intraday and imbalance markets. One could argue that in a sense the day-ahead auction is a forward market anticipating the real-time (balancing) price, so the latter would be the "true" spot price. To avoid confusion, instead of writing about "spot" prices, I refer to day-ahead, intraday and imbalance prices.

## 2.3. Price volatility vs. variability

Electricity prices fluctuate along two distinct time dimensions: trading time and delivery time. At one moment in time, different electricity products (i.e., different delivery periods) trade at different prices. In addition, as time passes, the price of each of them may change. Each is a separate phenomenon that can exist independently of the other, which I call variability and volatility (Figure 2).

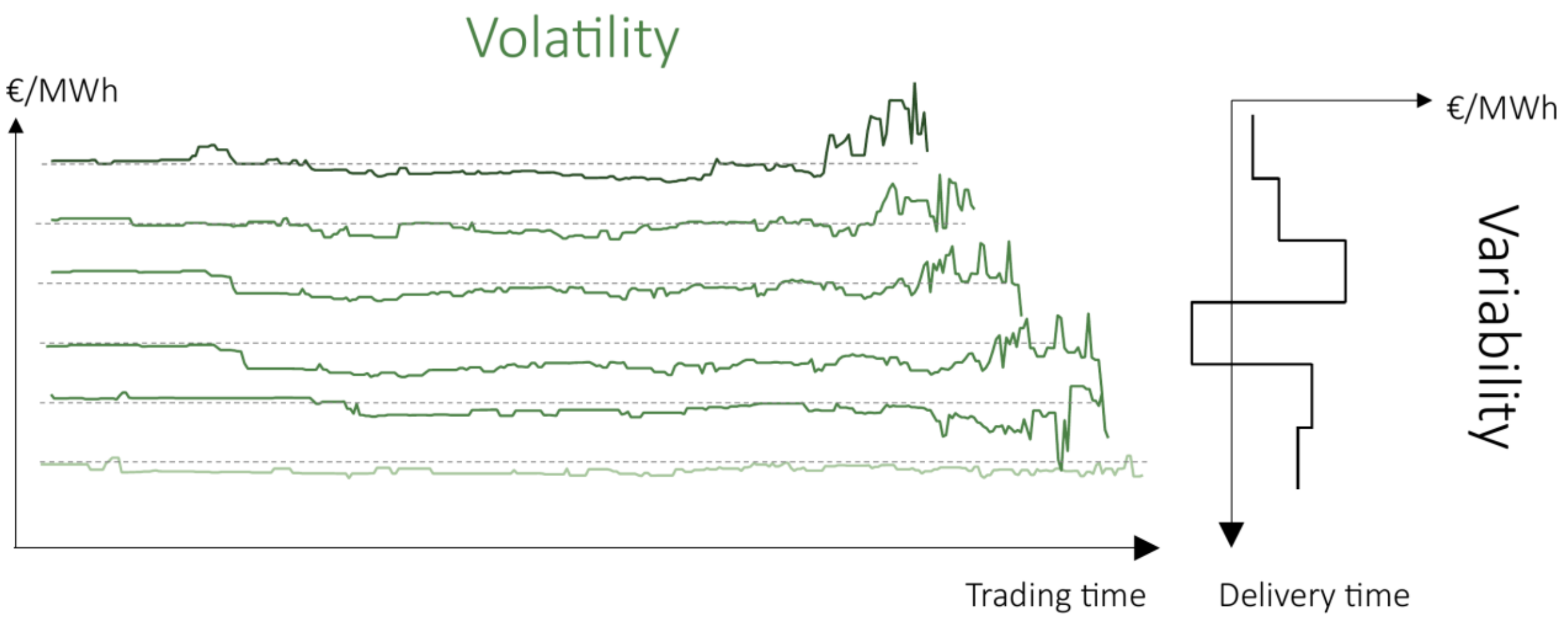

Figure 2. Volatility and variability: two dimensions of electricity price variation

Prices of different electricity goods differ, i.e., the market-clearing day-ahead price is different for each quarter-hour. These differences would persist even in the case of perfect foresight. They are caused not by new information emerging, but by (predictable) differences in demand and supply conditions over time in conjunction with non-storability of electricity. I call this "price variability". Variability can be calculated from historical day-ahead prices, a time series of prices of different electricity goods. The main drivers of price variability are fluctuations of demand and the generation potential of wind and solar energy. Further predictable drivers are planned outages, heat-led combined heat and power, and seasonal hydro inflows. Such variability of quantities causes variability of prices. Hence variability requires three conditions: variability in quantities, non-storability and an upward-sloping supply curve. It occurs under perfect foresight; it does not require uncertainty. If electricity were non-storable but market parties had perfect foresight, prices would differ between delivery periods, but the price of each delivery period would not change over trading time: variability without volatility.

The price of any electricity good changes over trading time, between the day-ahead auction and realization (imbalance settlement). This is "volatility" as known in the finance literature. It is a consequence of uncertainty and new information arriving; non-storability is no precondition for it to occur. If electricity could be stored at no cost but the future were uncertain, arbitrage would equalize prices across delivery periods, but prices would move with news: volatility without variability (the prices of all electricity products traded at that moment would move in sync). Volatility can be observed as realized volatility from past trades, a time series of prices for the same electricity good traded at different times to maturity. To my knowledge, no derivative markets for intraday or imbalance prices exist, so implied volatility cannot be derived from option prices (implied volatility can be calculated for forward markets, but those are bundles of quarter-hours).

Table 1. Volatility vs. variability

| | Volatility | Variability |
|---|---|---|
| Time dimension | Trading time (time to maturity $\tau$) | Delivery time (market time unit $t$) |
| Necessary condition | New information arriving (realized) / uncertainty about the future (implied) | Non-storability in conjunction with differences in fundamentals (demand/supply) |
| Under perfect foresight | No | Yes (variability only) |
| With costless storage | Yes (volatility only) | No |
| Realization | Price path | Price shape, price profile |
| Measure | Dispersion of the price path across trading time | Dispersion of the price shape across delivery periods |

I denote by $p(\tau,t)$ the price at trading time $\tau$ for delivery period $t = 1,\dots,T$. The day-ahead price is $p(\tau_{DA},t)$, the imbalance price is $p(\tau_{IB},t)$. Volatility is the dispersion of $p(\tau,t)$ along $\tau$ for a fixed $t$; variability is the dispersion of $p(\tau,t)$ across $t$ for a fixed market stage $\tau$, usually the day-ahead auction (Table 1).

In principle, delivery time and trading time are both continuous. Market time units discretize delivery time, auctions discretize trading time. Discretization simplifies processes, reduces transaction costs and pools liquidity; in part, it also reflects convention and tradition.

## 2.4. Volatility in electricity forward markets

This paper is about volatility and variability in day-ahead, intraday and imbalance markets, where electricity is traded at quarter-hourly granularity. However, volatility is a concept well known and much studied in forward markets of commodities, including electricity. This section briefly refers to this field of study and explains how it differs from volatility and variability in short-term markets.

Consider the most common electricity forward contract in European markets: a financial front-year base forward. This is a financial derivative, whose underlying consists of 35,040 quarter-hourly day-ahead prices, each weighted equally. It corresponds to the delivery of 1 MW of power, supplied constantly throughout the next calendar year.

These forward contracts are traded for several years before delivery (Figure 3). During that time, like any other derivative, they exhibit volatility. Daily settlement prices can be used to calculate realized volatility, and observed prices of options on these forward contracts can be used to determine implied volatility. Such volatility matters in many respects, not least for the variation margins that traders are asked to deposit with clearing houses when holding short or long positions in exchange-traded power forwards (ECC 2026). In these aspects, electricity is no different from other commodities such as oil, gas or minerals. But in the case of electricity, this is not the end of the story.

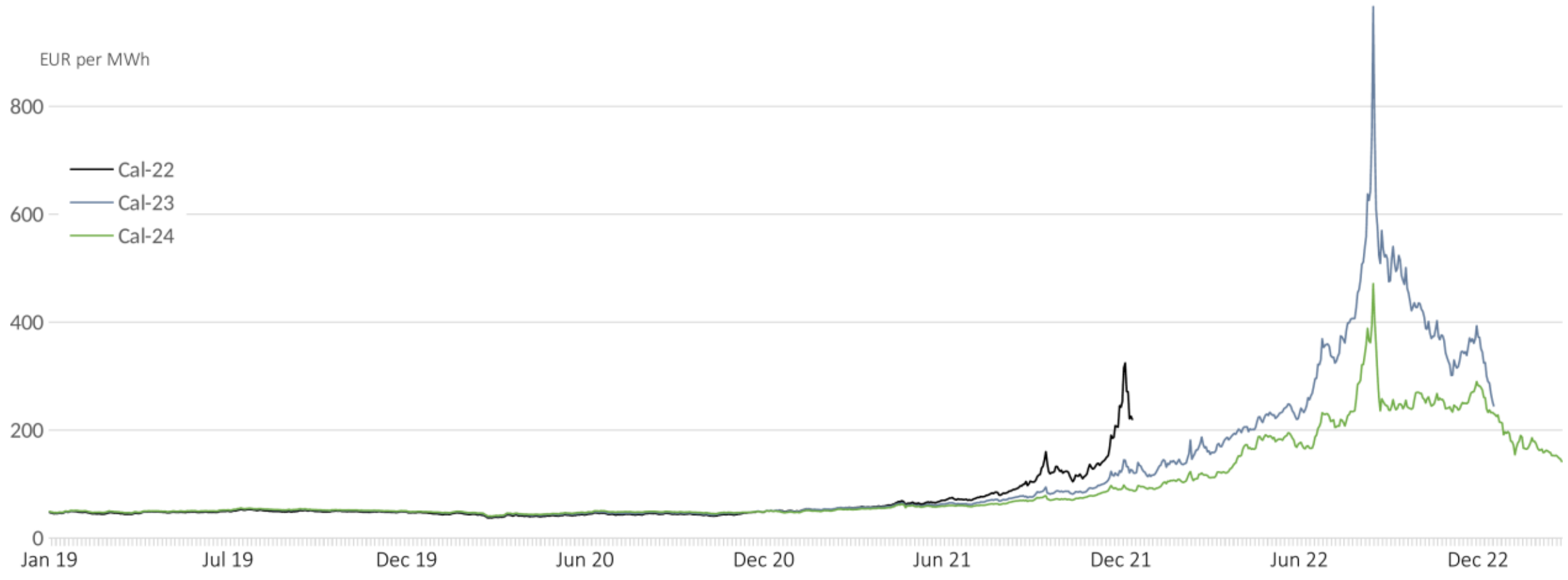

Figure 3. Electricity forward prices change over trading time, 2019–2023

The matter is significantly complicated by the fact that individual electricity goods (quarter-hourly products) are *not* traded on forward markets. Instead, it is common to bundle electricity forward contracts for longer periods such as months, quarters and years. A base year forward contract really is a strip of 35,040 quarter-hourly contracts. The fact that native electricity goods are bundled was even true for the time horizon commonly referred to as the "spot market": the European day-ahead auction used to trade hourly electricity contracts, each of which really is a bundle of quarter-hourly electricity contracts (most exchanges introduced quarter-hourly products only in 2025).

Because a strip averages over its components, its volatility is lower than the volatility of the individual contracts, so that forward contracts with shorter delivery periods like weeks or months are more volatile (Benth and Koekebakker 2008; Kiesel et al. 2009; Benth and Schmeck 2014; Latini et al. 2019). The volatility of quarter-hourly electricity products cannot be recovered from the volatility of strips that bundle thousands of such products. The shape within a strip is not traded. Utilities and analysts model it with hourly price forward curves (Fleten and Lemming 2003; Caldana et al. 2017). The interplay of strips and volatility in forward markets is not the topic of this paper.

Variability should be confused neither with the term structure of volatility (the volatility of forward contracts with different maturities) nor with calendar spread volatility (the volatility of the spread between two forward contracts with different maturities).

# 3. Literature review

## 3.1. Price variability

Twenty-seven of the papers I reviewed study the dispersion of electricity prices across delivery periods, i.e., what I call variability (Table 2). Almost all studies refer to this as "volatility". A few use "variance" (Woo et al. 2011; Wozabal et al. 2016; Johnson and Oliver 2019; Schöniger and Morawetz 2022); Maciejowska (2020) uses both "variability" and "volatility". Some authors explicitly equate what they measure with price uncertainty (Ciarreta et al. 2020; Pavlík et al. 2025). Huisman et al. (2007) argue that the hours of a day-ahead auction are better treated as a cross-section of simultaneously determined prices than as a time series. Baule and Naumann (2021) are, to my knowledge, the only authors who remark on the conflation of the two concepts. Some publications use the term "price variability" for price changes over trading time (Newbery and Stiglitz 1981; Deaton and Laroque 1992), i.e., what this paper calls volatility.

These studies measure the dispersion of observed prices (day-ahead or real-time/imbalance prices) or the outcome of simulations.

Studies measure price variability in four major ways.

- The first is the variance or standard deviation of prices within a day or a week: of hourly day-ahead prices (Wozabal et al. 2016; Rintamäki et al. 2017; Schöniger and Morawetz 2022; Pavlík et al. 2025), of hourly real-time prices (Zareipour et al. 2007; Werner 2014), of 15-min real-time prices (Woo et al. 2011) or of 5-min prices within the hour (Saretto et al. 2024). Schischke and Rathgeber (2026) sum squared changes between consecutive hours.

- The second is the dispersion of daily prices over longer periods, such as the standard deviation of daily prices within a month (Cevik and Ninomiya 2023) or the variance of daily prices within a quarter (Johnson and Oliver 2019).
- The third is the conditional variance of daily or hourly prices in GARCH-type models (Ketterer 2014; Kyritsis et al. 2017; Do et al. 2019; Ciarreta et al. 2020; Mwampashi et al. 2021) or stochastic volatility models (Gianfreda et al. 2023; Kloster and Benth 2026). Related work models hourly day-ahead prices as mean-reverting jump-diffusion or regime-switching processes, mostly for short-term price forecasting (de Jong and Huisman 2002; Huisman and Mahieu 2003; Mount et al. 2006; Weron 2006), and characterizes price spikes as a distinctive feature of electricity prices (Geman and Roncoroni 2006).
- The fourth uses quantiles: inter-quantile ranges (Maciejowska 2020), quantile regressions (Sapio 2019; Do et al. 2019; Owolabi et al. 2022), the band between the 5th and 95th percentile of hourly prices (Seel et al. 2018) or the frequency of extreme prices (Rai and Nunn 2020). Simulation studies apply such measures to modeled prices (Green and Vasilakos 2010; Milstein and Tishler 2011; Brancucci Martinez-Anido et al. 2016; Seel et al. 2018).

Most empirical studies find that wind power increases price variability, while the effect of solar power differs between systems. Wind raises variability in Texas (Woo et al. 2011), Germany (Ketterer 2014; Kyritsis et al. 2017), Italy (Sapio 2019) and Australia (Mwampashi et al. 2021), but lowers daily variability in Denmark (Rintamäki et al. 2017). In German data, Ketterer (2014) finds that wind power lowers the price level but raises the variance of log day-ahead prices in a GARCH model. Solar power lowered variability in Germany in the early 2010s (Kyritsis et al. 2017; Rintamäki et al. 2017) and raises it in recent European data (Pavlík et al. 2025; Schischke and Rathgeber 2026). Several studies find non-monotonic effects: small shares of renewables reduce price variance, large shares increase it (Wozabal et al. 2016; Schöniger and Morawetz 2022). The sign of the wind effect depends on demand (Maciejowska 2020). In six US markets, higher shares of wind and solar power coincide with lower variability in five (Owolabi et al. 2022). Simulations without uncertainty also yield higher variability with more wind or solar power (Green and Vasilakos 2010; Milstein and Tishler 2011; Seel et al. 2018); in New England, the effect is larger for 5-min than for hourly price changes (Brancucci Martinez-Anido et al. 2016). Other factors matter as much as renewables: interconnection (Sapio 2019; Schöniger and Morawetz 2022), flexible gas-fired capacity (Werner 2014; Saretto et al. 2024), conventional generation (Schischke and Rathgeber 2026) and regulation (Ciarreta et al. 2020). Long-run evidence across countries is scarce: Johnson and Oliver (2019) cover 19 countries in 2000–2011, Cevik and Ninomiya (2023) 24 European countries in 2014–2021. Mercier et al. (2023) study the value of storage arbitrage on day-ahead markets, which depends on price variability, across Europe.

Table 2. Studies on electricity price variability

| Study | Case | Prices | Metric | Term used |
|---|---|---|---|---|
| Zareipour et al. (2007) | Ontario and neighboring markets | Real time, hourly | Std. dev. of log price returns (historical volatility), price velocity | Volatility |
| Green and Vasilakos (2010) | GB | Simulation | No formal metric (descriptive) | Volatility |
| Jacobsen and Zvingilaite (2010) | DK | Day-ahead | No formal metric (descriptive) | Volatility, variation |
| Milstein and Tishler (2011) | Israel | Simulation | Dispersion of daily prices | Volatility |
| Woo et al. (2011) | ERCOT, 2007–2010 | Real time, 15-min | Variance | Variance |
| Ketterer (2014) | DE, 2006–2011 | Day-ahead | Conditional variance (GARCH) of daily prices | Volatility |
| Werner (2014) | ISO-NE, 2005–2011 | Real time | Std. dev. of hourly prices within a day | Volatility |
| Brancucci Martinez-Anido et al. (2016) | ISO-NE | Simulation | 5-min and hourly price changes | Volatility |
| Wozabal et al. (2016) | DE/AT, 2007–2013 | Day-ahead | Variance of hourly prices within a day | Variance |
| Kyritsis et al. (2017) | DE, 2010–2015 | Day-ahead | Conditional variance (GARCH-in-mean) of daily prices | Volatility |
| Rintamäki et al. (2017) | DK, DE, 2010–2014 | Day-ahead | Std. dev. of hourly prices within a day/week | Volatility |
| Seel et al. (2018) | 4 US markets, 2030 | Simulation | 5th–95th percentile band, coefficient of variation | Volatility |
| Do et al. (2019) | DE, 2015–2018 | Day-ahead | Conditional variance (EGARCH), conditional quantiles | Volatility |
| Johnson and Oliver (2019) | 19 countries, 2000–2011 | – | Quarterly variance of daily prices | Variance |
| Sapio (2019) | IT, 2006–2015 | Day-ahead | Conditional quantiles (quantile regression) | Volatility |
| Ciarreta et al. (2020) | ES, 2002–2017 | Day-ahead | Conditional variance with structural breaks | Volatility |
| Maciejowska (2020) | DE | Day-ahead | Interquantile range | Variability, volatility |
| Rai and Nunn (2020) | South Australia | Real time, 30-min | Std. dev., frequency of extreme prices | Volatility |
| Mwampashi et al. (2021) | Australia (NEM) | Real time, daily | Conditional variance (EGARCH) | Volatility |
| Owolabi et al. (2022) | 6 US markets | Real time, hourly | Conditional quantiles (quantile regression), scale of skew-t distribution | Volatility |
| Schöniger and Morawetz (2022) | 9 countries, 2015–2019 | Day-ahead | Variance of hourly prices within a day | Variance |
| Cevik and Ninomiya (2023) | 24 European countries, 2014–2021 | – | Std. dev. of daily prices within a month | Volatility |
| Gianfreda et al. (2023) | DE, IT, DK, 2016–2019 | Day-ahead | Stochastic volatility (time-varying variance) of hourly prices | Volatility |
| Saretto et al. (2024) | PJM, 2014–2016 | Real time, 5-min | Range and std. dev. of 5-min prices within the hour | Volatility |
| Pavlík et al. (2025) | 6 European countries, 2015–2025 | Day-ahead | Std. dev. of hourly or 15-min prices within a day | Volatility |
| Kloster and Benth (2026) | 3 European bidding zones | Day-ahead | Integrated variance of a latent price process | Volatility |
| Schischke and Rathgeber (2026) | 13 European markets, 2015–2023 | Day-ahead | Sum of squared hourly price changes per day (realized variance) | Volatility |

## 3.2. Price volatility

Most studies of price volatility on intraday markets use “volatility” in the sense of this paper, i.e., for price changes of the same contract across trading time (Ciarreta et al. 2017; Baule and Naumann 2021; Birkeland et al. 2024; Richard 2025). Others do not name a dispersion concept and model price changes or price differences (Kremer et al. 2020, 2021; Kiesel and Paraschiv 2017); Woo et al. (2016) speak of “price divergence” between day-ahead and real-time prices (Table 3).

Studies of price volatility across trading time use intraday and real-time market data. Some measure the dispersion of trade prices of a given contract during its trading session: the standard deviation of price changes between consecutive trades and related range measures (Baule and Naumann 2021), the realized variance of high-frequency trade prices (Ciarreta et al. 2017; Birkeland et al. 2024) or the standard deviation of trade prices of a product during a session (Richard 2025). Others model the changes of a contract’s price between trades as a function of forecast updates (Kremer et al. 2020, 2021), the distribution of intraday returns (Hirsch and Ziel 2024), intraday price levels relative to the day-ahead price (Hagemann 2015; Pape et al. 2016; Kiesel and Paraschiv 2017; Kulakov and Ziel 2021) or the difference between day-ahead and real-time prices (Woo et al. 2016). None of the studies I reviewed uses imbalance prices.

Forecast errors and forecast updates of wind and solar power drive intraday prices and their changes (Hagemann 2015; Kiesel and Paraschiv 2017; Kremer et al. 2020, 2021; Kulakov and Ziel 2021). The response is asymmetric and stronger when the merit order is steep (Kiesel and Paraschiv 2017; Kremer et al. 2021). Wind power increases the dispersion of trade prices of hourly contracts in Germany (Baule and Naumann 2021). Additional wind and solar capacity changes intraday price volatility in a non-linear way (Kulakov and Ziel 2021). In the Netherlands, forecast errors of load and renewables do not increase realized volatility (Birkeland et al. 2024). Richard (2025) relates traded volume, price level and volatility in the French intraday market to balancing needs, short-term flexibility and forecast errors of renewables in France and abroad. In California, the difference between day-ahead and real-time prices partly depends on the day-ahead forecast errors for load and renewables (Woo et al. 2016). None of these studies traces volatility over the years of renewable expansion.

Table 3. Studies on electricity price volatility

| Study | Case | Prices | Measure | Term used |
|---|---|---|---|---|
| Hagemann (2015) | DE, 2010–2011 | Continuous intraday | No dispersion metric (regression of prices on forecast errors) | – |
| Pape et al. (2016) | DE, 2012–2013 | Day-ahead and intraday | No dispersion metric (fundamental model of price levels) | Price variations |
| Woo et al. (2016) | CAISO, 2012–2015 | Day-ahead and real time | Real-time minus day-ahead price | Price divergence |
| Ciarreta et al. (2017) | DE/AT | Continuous intraday | Realized variance of trade prices | Volatility |
| Kiesel and Paraschiv (2017) | DE | Continuous intraday, 15-min | No dispersion metric (regression of prices on forecast errors) | – |
| Kremer et al. (2020) | DE, 2015 | Continuous intraday | No dispersion metric (regression of price changes between trades) | – |
| Kremer et al. (2021) | DE, 2015 | Continuous intraday | No dispersion metric (regression of price changes between trades) | – |
| Baule and Naumann (2021) | DE, 2015–2018 | Continuous intraday | Std. dev. of price changes, ranges | Volatility |
| Kulakov and Ziel (2021) | DE | Continuous intraday | No dispersion metric (intraday price from shifted auction curves) | Volatility |
| Birkeland et al. (2024) | NL | Continuous intraday | Realized variance of trade prices | Volatility |
| Hirsch and Ziel (2024) | DE | Continuous intraday | Std. dev., median absolute deviation and interquartile range of price changes | Volatility |
| Richard (2025) | FR, 2024 | Continuous intraday | Std. dev. of trade prices per session | Volatility |

## 3.3. Renewables system integration

A large body of literature studies the integration costs and the system value of wind and solar power (Grubb 1991; Hirst and Hild 2004; DeCarolis and Keith 2005; Holttinen 2005; Fripp and Wiser 2008; Lamont 2008; Milligan and Kirby 2009; Mills and Wiser 2010, 2012; Joskow 2011; Milligan et al. 2011; Katzenstein and Apt 2012; Lannoye et al. 2012; Ueckerdt et al. 2013; Hirth 2013; Hirth et al. 2015; Hirth et al. 2016; Gowrisankaran et al. 2016; Wiser et al. 2017; Joos and Staffell 2018; López Prol et al. 2020; Heptonstall and Gross 2021). In this branch of the literature, "variability" denotes the fluctuation of generation potentials over time, which would persist under perfect foresight. This is consistent with my use of "price variability" in the present paper. In the same way, Wiser et al. (2017) use "price variability" for the effect of wind and solar energy on wholesale prices across delivery periods.

Uncertainty denotes forecast errors (Ela et al. 2011; Milligan et al. 2011). Variability of generation potentials and demand causes price variability; uncertainty causes price volatility. Hirth et al. (2016) describe electricity as heterogeneous along three dimensions: time, space and lead time. Time corresponds to the delivery period $t$, lead time to the trading time $\tau$.

The system integration literature studies the drivers of price variability and volatility rather than the price phenomena themselves. Wind and solar power are a source of both. The variability of their generation potentials shapes residual load across delivery periods and thereby the day-ahead price shape; it lowers the market value of wind and solar power as their share grows (Hirth 2013; López Prol et al. 2020). Their forecast errors are a source of uncertainty that is resolved in intraday trading and balancing (Hirth and Ziegenhagen 2015; Joos and Staffell 2018; Koch and Hirth 2019) and that moves imbalance volumes and prices (Goodarzi et al. 2019). In the terms of integration costs, profile costs correspond to variability, balancing costs to volatility (Hirth et al. 2015; Heptonstall and Gross 2021).

## 3.4. Volatility in finance

In the field of finance, volatility is usually understood as a property of a time series of prices of a financial asset such as a stock or bond. It is often measured as the standard deviation of logarithmic returns. Both realized and implied volatility are of interest.

Actual historical (or realized) volatility can be calculated from observed historical settlement prices of the financial asset itself. It is a measure of how strongly an asset price has fluctuated in the past. Realized volatility exists because new information became available at different moments in time, changing the value of the asset. For example, the news of an oil field being attacked causes the oil price to jump. If all relevant information had been available at the beginning of the observation period, returns would have remained stable and realized volatility zero.

Expected (or implied) volatility of a financial asset is usually inferred from the price of options on that asset. Option pricing theory, dating back to Black and Scholes (1973) and Merton (1973), shows that volatility is intimately related to the fair price of an option: a more volatile asset price implies a larger chance that a call or put option ends up in the money and hence a larger fair option value (Figure 4). Implied volatility can be understood as uncertainty about future price movements, which, in the case of nonlinear financial derivatives such as call and put options, increases the value of such options. The larger the implied volatility, the higher the extrinsic (or time) value of an option, or rather: a high extrinsic value is an expression of high implied volatility. Implied volatility exists because the future is uncertain – under perfect foresight, implied volatility, and hence the extrinsic value of an option, would be zero.

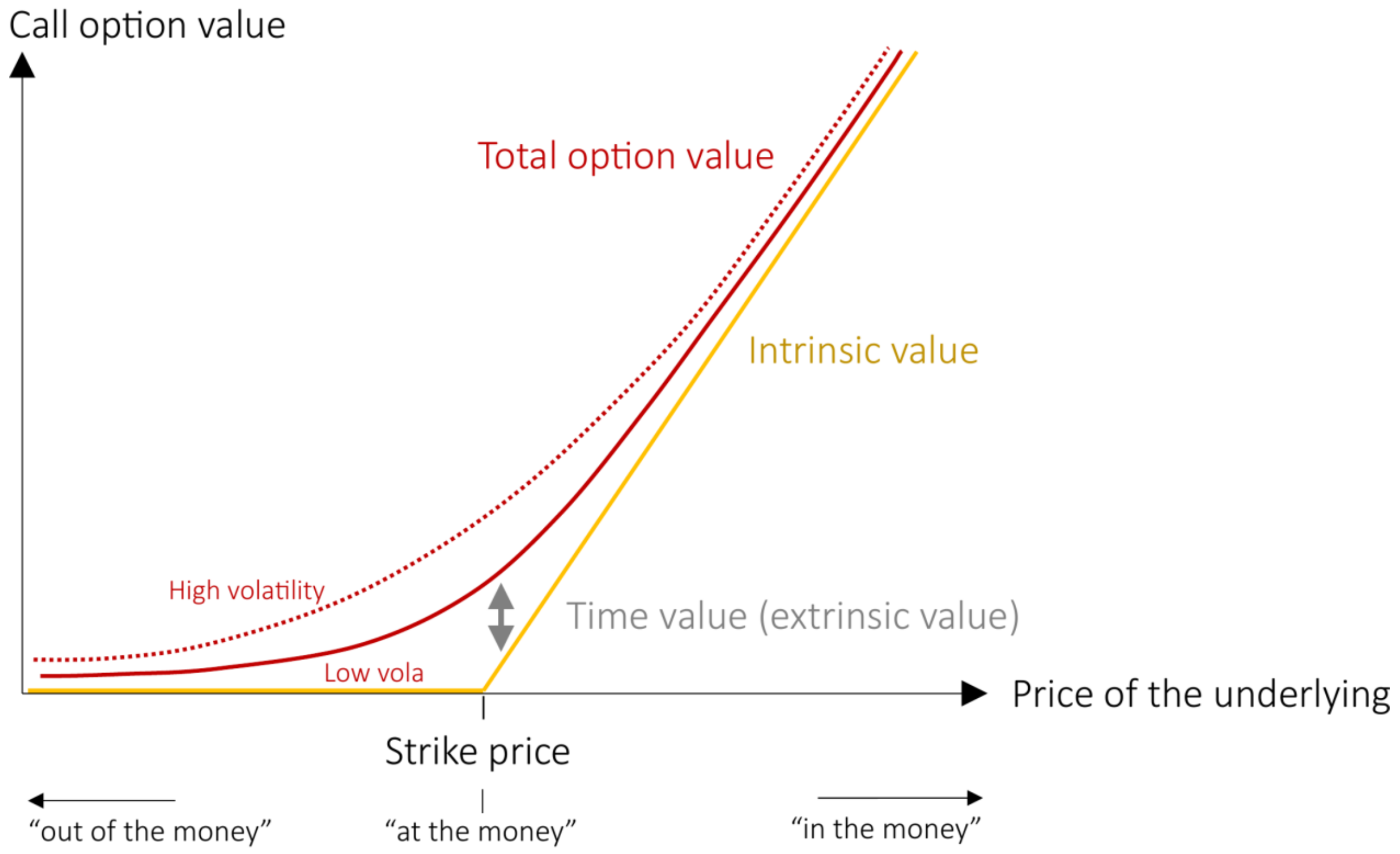

Figure 4. Fair option value for different levels of implied volatility

In the finance literature, volatility is a property of the price of one asset across trading time. This paper uses the term in the same sense for electricity: volatility is the dispersion of the price of a given delivery period across trading time. Price differences across delivery periods are not volatility, because they are prices of different goods.

# 4. An analytical model of volatility and variability

This section formalizes the two concepts of price variability and volatility, identifies their drivers and derives hypotheses for Section 6. A linear merit order, price-inelastic load and rational expectations form the base case. Storage, a convex merit order and a risk premium are extensions. Table 4 lists the symbols.

Table 4. Symbols

| Symbol | Meaning | Unit |
|---|---|---|
| $\tau$ | Trading time; $\tau_{DA}$ day-ahead auction, $\tau_{ID}$ intraday gate closure, $\tau_{IB}$ settlement | – |
| $t$ | Delivery period (market time unit); $T$ last delivery period of the year; $N$ periods per day | – |
| $p(\tau, t)$ | Price at trading time $\tau$ for delivery period $t$; $p(\tau_{IB}, t)$ realized price | €/MWh |
| $r(t)$ | Spread $p(\tau_{IB}, t) - p(\tau_{DA}, t)$ | €/MWh |
| $L(t)$ | Load (price-inelastic) | GW |
| $K_W, K_S$ | Installed wind and solar capacity | GW |
| $w(t), s(t)$ | Wind and solar capacity factors | – |
| $R(t)$ | Residual load | GW |
| $a, b$ | Intercept and slope of the linear merit order; $b_{ID}$, $b_{IB}$ slopes at the intraday and imbalance settlement stage | €/MWh; €/MWh per GW |
| $\hat{R}(t)$ | Expectation at $\tau_{DA}$ (day-ahead forecast); hats on $L, w, s$ likewise | GW |
| $\bar{R}$ | Mean over the delivery periods of a day; bars on $\varepsilon, p$ likewise | GW |
| $\varepsilon(t)$ | Forecast error of residual load, $R(t) - \hat{R}(t)$; components $\varepsilon_L$, $\varepsilon_w$, $\varepsilon_s$, $\varepsilon_G$ (load, wind, solar, plant outages) | GW |
| $\sigma_L, \sigma_w, \sigma_s, \sigma_G$ | Standard deviations of the error components ($\sigma_w$, $\sigma_s$ per unit of capacity) | GW; – |
| $\sigma_{DA}, \sigma_{ID}$ | Standard deviation of the residual-load forecast error at the day-ahead auction and at gate closure | GW |
| $\mathrm{Var}_t, \mathrm{Cov}_t$ | Variance and covariance across delivery periods | – |
| $\mathrm{E}_\tau, \mathrm{Var}$ | Expectation and variance over uncertainty, conditional on information at $\tau$ | – |
| $K^*$ | Solar capacity minimizing within-day variability | GW |
| $x(\tau, t)$ | Storage net discharge for $t$ scheduled at $\tau$ | GW |
| $P, E$ | Storage power and energy capacity | GW, GWh |
| $e(t)$ | State of charge | GWh |
| $\eta(\tau)$ | Storage price response at trading stage $\tau$ | GW per €/MWh |
| $\tilde{b}(\tau)$ | Effective merit-order slope with storage | €/MWh per GW |
| $\delta$ | Storage cycle cost | €/MWh |
| $D$ | Residual load spread between two delivery periods, $R(2) - R(1)$ | GW |
| $V(\cdot)$ | Value of the storage schedule as a function of the price vector | € |
| $c(\cdot)$ | Convex merit order (extension) | €/MWh |
| $\pi(\cdot)$ | Risk premium (extension) | €/MWh |

## 4.1. Price formation

Delivery periods $t = 1, \dots, N$ form one day; trading takes place at the day-ahead auction $\tau_{DA}$, at intraday gate closure $\tau_{ID}$ and at settlement $\tau_{IB}$. Load $L(t)$ is price-inelastic. Wind and solar generation are the product of installed capacity and capacity factor. Residual load is

$$R(t) = L(t) - K_W \cdot w(t) - K_S \cdot s(t) \tag{1}$$

Thermal plants supply the residual load along a linear merit order with intercept $a$ and slope $b$; both scale with fuel and carbon prices. The realized price is

$$p(\tau_{IB}, t) = a + b \cdot R(t) \tag{2}$$

At the day-ahead auction, market parties hold forecasts of load and capacity factors. Realized residual load deviates from its forecast by the error $\varepsilon(t)$, which collects the load forecast error, the wind and solar forecast errors and plant outages:

$$R(t) = \hat{R}(t) + \varepsilon(t), \quad \varepsilon(t) = \varepsilon_L(t) - K_W \cdot \varepsilon_w(t) - K_S \cdot \varepsilon_s(t) + \varepsilon_G(t) \tag{3}$$

Prices are conditional expectations; there is no risk premium. With a linear merit order, the day-ahead price equals the price at the forecast residual load. The imbalance-DA spread (Table 4) is proportional to the forecast error:

$$p(\tau_{DA}, t) = \mathrm{E}_{\mathrm{DA}}[p(\tau_{IB}, t)] = a + b \cdot \hat{R}(t) \tag{4}$$

$$r(t) = b \cdot \varepsilon(t) \tag{5}$$

## 4.2. Variability

Within-day variability of day-ahead prices is the slope times the dispersion of forecast residual load across delivery periods. The dispersion decomposes into the profiles of load, wind and solar and their covariances:

$$\mathrm{Var}_t[p(\tau_{DA}, t)] = b^2 \cdot \mathrm{Var}_t[\hat{R}(t)] \tag{6}$$

$$\mathrm{Var}_t[\hat{R}] = \mathrm{Var}_t[\hat{L}] + {K_W}^2 \cdot \mathrm{Var}_t[\hat{w}] + {K_S}^2 \cdot \mathrm{Var}_t[\hat{s}] - 2 \cdot K_W \cdot \mathrm{Cov}_t[\hat{L}, \hat{w}] - 2 \cdot K_S \cdot \mathrm{Cov}_t[\hat{L}, \hat{s}] + 2 \cdot K_W \cdot K_S \cdot \mathrm{Cov}_t[\hat{w}, \hat{s}] \tag{7}$$

The solar term is not monotonic. Solar generation is positively correlated with load within the day, so a small solar fleet flattens residual load. Additional capacity raises within-day variability beyond the capacity

$$K^* = \frac{\mathrm{Cov}_t[\hat{L}, \hat{s}] - K_W \cdot \mathrm{Cov}_t[\hat{w}, \hat{s}]}{\mathrm{Var}_t[\hat{s}]} \tag{8}$$

The within-day profile of wind is weak, so its contribution to (7) is small. Wind dominates between-day variability, the dispersion of daily mean prices across days. Equations (6) and (7) hold for this measure with variances and covariances of daily means.

Under perfect foresight, $\varepsilon = 0$ and $r = 0$: variability only, as in Table 1. Variability requires $b > 0$ and non-storability; Section 4.4 shows how storage removes it.

## 4.3. Volatility

With independent error components, volatility is the slope times the variance of the forecast error:

$$\mathrm{Var}[r(t)] = b^2 \cdot (\sigma_L^2 + {K_W}^2 \cdot \sigma_w^2 + {K_S}^2 \cdot \sigma_s^2 + \sigma_G^2) \tag{9}$$

Wind and solar enter monotonically and in proportion to installed capacity; there is no counterpart to $K^*$. Solar forecast errors are confined to daylight hours.

Part of the forecast error is resolved before gate closure; the remainder is settled at the imbalance price. Each stage has its own slope:

$$\mathrm{Var}[p(\tau_{IB}, t) - p(\tau_{DA}, t)] = b_{ID}^2 \cdot (\sigma_{DA}^2 - \sigma_{ID}^2) + b_{IB}^2 \cdot \sigma_{ID}^2 \tag{10}$$

$b_{ID}$ exceeds $b$ because fewer plants respond at short notice. $b_{IB}$ is set by the settlement formula (Section 2.2). The settlement formula therefore affects volatility but not variability.

Fuel and carbon prices scale variability and volatility proportionally through $b$. The ratio of volatility to variability, $\frac{b_{ID}}{b} \cdot \frac{\mathrm{sd}[\varepsilon]}{\mathrm{sd}_t[\hat{R}]}$, does not depend on fuel prices.

Since $\mathrm{E_{DA}}[\varepsilon(t) \mid \hat{R}(t)] = 0$, the covariance term in (25) (Section 5.2) is zero, also for a convex merit order. A negative covariance requires a state-dependent risk premium or flexibility that responds to the day-ahead shape only after the day-ahead auction.

## 4.4. Storage

A storage fleet with power $P$ and energy $E$ chooses a schedule $x(\tau, t)$ at each trading stage and trades it again at later stages, subject to the power and energy constraints. Storage enters the price as additional supply:

$$-P \leq x(\tau_{IB}, t) \leq P, \quad 0 \leq e(t) \leq E, \quad e(t) = e(t-1) - x(\tau_{IB}, t) \tag{11}$$

$$p(\tau_{IB}, t) = a + b \cdot [R(t) - x(\tau_{IB}, t)] \tag{12}$$

Revenue equals the value of the final schedule minus the spread on the day-ahead position:

$$\Pi = \textstyle\sum_t p(\tau_{IB}, t) \cdot x(\tau_{IB}, t) - \sum_t r(t) \cdot x(\tau_{DA}, t) \tag{13}$$

For a risk-neutral price taker, $\mathrm{E_{DA}}[\Pi]$ does not depend on the day-ahead position, because the spread has zero mean. The value of flexibility is $\mathrm{E_{DA}}[V(p(\tau_{IB},\cdot))] - V(p(\tau_{DA},\cdot))$, where $V$ is the value of the schedule as a function of prices. $V$ is convex, so the value of flexibility increases with volatility.

Market clearing pins down the day-ahead position of the fleet. The intraday redispatch $x(\tau_{IB}, t) - x(\tau_{DA}, t)$ absorbs the surprise:

$$x(\tau_{DA}, t) = \mathrm{E_{DA}}[x(\tau_{IB}, t)] \tag{14}$$

If the fleet is marginal in every period and energy neutral within the day, its response is linear in the deviation of the price from the daily mean. Prices then follow residual load with the effective slope $\tilde{b}$. Volatility splits into a within-day component, which storage dampens, and a daily mean component, which passes through:

$$x(\tau,t) = \eta(\tau) \cdot [p(\tau,t) - \bar{p}(\tau)], \quad \sum_t x(\tau,t) = 0 \tag{15}$$

$$p(\tau,t) - \bar{p}(\tau) = \tilde{b}(\tau) \cdot [R(t) - \bar{R}], \quad \tilde{b}(\tau) = \frac{b}{1 + b \cdot \eta(\tau)} \tag{16}$$

$$\mathrm{Var}[r(t)] = \tilde{b}_{ID}^2 \cdot \mathrm{Var}[\varepsilon(t) - \bar{\varepsilon}] + b^2 \cdot \mathrm{Var}[\bar{\varepsilon}] \tag{17}$$

Cycles across delivery periods flatten the shape; redispatch across trading time absorbs the within-day part of the forecast error. More generally, storage with horizon $H$ dampens fluctuations shorter than $H$ on both axes. Unlimited storage removes variability and the volatility caused by forecast errors; only permanent shocks remain, the storable case of Table 1.

With two delivery periods, $E = P$, cycle cost $\delta$ and residual load spread $D = R(2) - R(1)$, a competitive fleet has three regimes:

$$x^* = \begin{cases} 0 & \text{for } b \cdot D \leq \delta \\ \frac{b \cdot D - \delta}{2 \cdot b} & \text{for } \delta < b \cdot D < \delta + 2 \cdot b \cdot P \\ P & \text{for } b \cdot D \geq \delta + 2 \cdot b \cdot P \end{cases} \tag{18}$$

The price spread is $b \cdot D$ when the fleet is idle, $\delta$ when it is marginal and $b \cdot D - 2 \cdot b \cdot P$ when it is saturated. The within-day error $\varepsilon(2) - \varepsilon(1)$ is fully absorbed only in the marginal regime. Dampening volatility requires headroom: a fleet saturated by the day-ahead shape passes shocks through. Variability and volatility compete for the same storage power.

## 4.5. Extensions

With $c(R)$ convex, the slope rises with residual load; both measures are higher in winter and in the evening. A floor segment with zero slope (curtailment, must-run) bounds prices from below. Solar beyond the floor raises residual load variance but not price variability. Forecast errors in floor hours have no price effect. The day-ahead price carries a convexity premium:

$$p(\tau_{DA},t) = \mathrm{E}_{\mathrm{DA}}[c(R(t))] \approx c(\hat{R}(t)) + ½ \cdot c''(\hat{R}(t)) \cdot \mathrm{Var}[\varepsilon(t)] \tag{19}$$

With $p(\tau_{DA},t) = \mathrm{E}_{\mathrm{DA}}[p(\tau_{IB},t)] + \pi(\hat{R}(t))$ and $\pi$ increasing in $\hat{R}$, the covariance term in (25) (Section 5.2) is negative.

With hourly day-ahead products and quarter-hourly delivery, the day-ahead price of the hourly product is the mean of four quarter-hourly expectations. The within-hour shape then appears in the spread: variability is measured as volatility. Section 5 therefore averages intraday prices to hours.

## 4.6. Implications for the empirical analysis

The model yields the following hypotheses for Section 6. Variability is explained by the within-day dispersion of forecast residual load times a fuel price proxy; the sign of solar depends on $K_S$ relative to $K^*$. Volatility is explained by the standard deviation of forecast errors times a fuel price proxy, in proportion to capacity rather than generation. The co-movement of the two measures (Figure 11) reflects the common factor $b$, not causation between them; the ratio varies only with $\mathrm{sd}[\varepsilon]/\mathrm{sd}_t[\hat{R}]$. The decomposition (Section 6.4) tests the benchmark $\mathrm{Cov}_t[p(\tau_{DA},t), r] = 0$; a regression of $r(t)$ on $p(\tau_{DA}, t) - \bar{p}(\tau_{DA})$ identifies a state-dependent premium. Storage dampens both measures, volatility only in the component within the storage horizon and in periods with headroom.

# 5. Data and measures

This section introduces the empirical data and the metrics used to calculate variability and volatility, my preferred metric being the trimmed and normalized standard deviation.

## 5.1. Data

Table 5. Data

| Series | Resolution | Period | Source |
|---|---|---|---|
| Day-ahead price (DE-AT-LU until September 2018, DE-LU thereafter) | Hourly; quarter-hourly from October 2025 | Jan 2001 – Sep 2026 | EPEX SPOT |
| Intraday auctions (opening auction/IDA1, IDA2, IDA3) | Quarter-hourly | Dec 2014 – Dec 2025 | EPEX SPOT |
| Continuous intraday: IDFULL, ID3, ID1 (volume-weighted average prices) | Hourly and quarter-hourly products | 2011 – Dec 2025 | EPEX SPOT |
| Continuous intraday: all transactions (hourly and quarter-hourly products) | Trade level (tick data) | Jan 2015 – Dec 2025 | EPEX SPOT |
| Imbalance price | Quarter-hourly | Jun 2010 – Oct 2026 | Amprion, Netztransparenz |
| Wind and solar generation | Quarter-hourly | 2015 – Sep 2026 | ENTSO-E Transparency Platform |
| 23 additional European bidding zones: day-ahead price, wind and solar generation, load | Hourly and quarter-hourly | 2015 – Sep 2026 | ENTSO-E Transparency Platform |
| Natural gas (TTF) | Daily | 2008 – 2023 | ICE Endex |
| EU allowance prices | Daily | 2008 – 2023 | EEX, ICE (EU allowances) |

Table 5 lists the data. Day-ahead prices for Germany cover January 2001 to early September 2026. Until 2004, I use the hourly auction prices of LPX and EEX as published by Energinet; from 2005 to 2025, EPEX SPOT data; for 2026, the ENTSO-E Transparency Platform. Where the sources overlap, they are identical, with one exception: on 26 June 2024, the European day-ahead market coupling was partially decoupled. The EPEX SPOT fallback auction cleared at prices of up to 2,326 €/MWh. I exclude this day from all measures. Intraday data comprise the quarter-hourly opening auction (from December 2014, succeeded by IDA1 in June 2024), the second and third intraday auctions (from June 2024) and the volume-weighted average prices of continuous trading for each hourly and quarter-hourly product: the full session (IDFULL), the last three hours (ID3) and the last hour before delivery (ID1). Quarter-hourly intraday prices are averaged to hours (Section 5.2).

The national German imbalance price reBAP has existed since June 2010; before, imbalance prices were set by each control area separately. I use the imbalance price from June 2010 onward. Its settlement formula changed repeatedly, notably in December 2012, May 2016, 2020 and June 2022. Imbalance settlement is quarter-hourly; I average it to hours, because the hourly day-ahead price contains no information on the split of an hour into quarter-hours.

Intraday trading was thin in the early years of the sample, and trading stages were added step by step: the opening auction in December 2014, the ID3 index in June 2015, the ID1 index in February 2018 and the second and third intraday auctions in June 2024 (Table 5). Comparisons across trading stages are therefore possible only from these dates. Actual wind and solar generation for Germany is taken from the ENTSO-E Transparency Platform.

For hourly and quarter-hourly products of continuous trading, I use all transactions from January 2015 to December 2025, about 900 million trades, including local and cross-border trades, without auction trades, self-trades and user-defined blocks (Table 5). Trades that deviate from the median price of their product by more than eight median absolute deviations and more than 50 €/MWh are dropped, less than 0.5 % of all trades.

For the comparison across Europe (Section 6.9), day-ahead prices, wind and solar generation and load of 23 further bidding zones are from the ENTSO-E Transparency Platform, January 2015 to early September 2026 (Table 5).

# 5.2. Dispersion metrics

The main measure for price dispersion on both dimensions (trading time and delivery time) is the trimmed, normalized standard deviation. Prices below the 1st and above the 99th percentile of each year are trimmed. The standard deviation is normalized using the annual average (base) day-ahead price.

Variability of a year is the trimmed standard deviation of hourly day-ahead prices, divided by the annual mean day-ahead price:

$$v_y = \frac{1}{\bar{p}_y} \cdot \mathrm{sd}_h\left(p_h^{\mathrm{DA}}\right) \tag{20}$$

Volatility at trading stage $s$ is the trimmed standard deviation of the hourly spread between the price at stage $s$ and the day-ahead price of the same delivery hour, divided by the annual mean day-ahead price. The standard deviation corrects for the mean spread, which differs from zero for the imbalance price in some years:

$$\sigma_y^s = \frac{1}{\bar{p}_y} \cdot \mathrm{sd}_h\left(p_h^s - p_h^{\mathrm{DA}}\right) \tag{21}$$

The variance of hourly prices around the annual mean decomposes exactly into a within-day component, a between-day component (daily means around the monthly mean $\bar{p}_{m(h)}$) and a between-month component:

$$\frac{1}{N_y} \cdot \sum_h \left(p_h - \bar{p}_y\right)^2 = \frac{1}{N_y} \cdot \sum_h \left(p_h - \bar{p}_{d(h)}\right)^2 + \frac{1}{N_y} \cdot \sum_h \left(\bar{p}_{d(h)} - \bar{p}_{m(h)}\right)^2 + \frac{1}{N_y} \cdot \sum_h \left(\bar{p}_{m(h)} - \bar{p}_y\right)^2 \tag{22}$$

Within-day variability of year $y$ is the root mean square deviation of hourly day-ahead prices $p_h$ from the mean price of the respective day $\bar{p}_{d(h)}$, divided by the annual mean price $\bar{p}_y$:

$$v_y^{wd} = \frac{1}{\bar{p}_y} \cdot \sqrt{\frac{1}{N_y} \cdot \sum_h \left(p_h - \bar{p}_{d(h)}\right)^2} \tag{23}$$

Between-day variability is the square root of the sum of the last two terms, divided by $\bar{p}_y$.

The imbalance price of delivery period $t$ equals the day-ahead price plus the imbalance-DA spread $r(t)$:

$$p(\tau_{IB}, t) = p(\tau_{\mathrm{DA}}, t) + r(t) \tag{24}$$

Taking the variance across delivery periods yields

$$\mathrm{Var}_t[p(\tau_{IB}, t)] = \mathrm{Var}_t[p(\tau_{\mathrm{DA}}, t)] + \mathrm{Var}_t[r] + 2 \cdot \mathrm{Cov}_t[p(\tau_{\mathrm{DA}}, t), r] \tag{25}$$

The variance of the imbalance price is the sum of day-ahead variability, the variance of the imbalance-DA spread and a covariance term. If prices are conditional expectations and carry no risk premium, the imbalance-DA spread is uncorrelated with the day-ahead price, so that the covariance term vanishes (Section 4.3). Under perfect foresight, $\mathrm{Var}[r] = 0$; with costless storage, day-ahead variability is zero. The imbalance price $p(\tau_{IB}, t)$ is the price at which every delivery period is ultimately settled; intraday prices are forward prices on it and reflect its expected value (Section 2.2). Because the imbalance price is set by a formula that has changed repeatedly, I apply the decomposition both to the imbalance price and to the ID3 index, the last liquid market price before delivery. Section 6.4 shows that the covariance term has been negative in Germany in every year since 2010.

The spread until the last trading hour is the sum of the increments $\Delta_k$ between consecutive trading stages (day-ahead auction, opening auction, full session, ID3, ID1):

$$\mathrm{Var}\left(p^{\mathrm{ID1}} - p^{\mathrm{DA}}\right) = \sum_k \mathrm{Var}\left(\Delta_k\right) + 2 \cdot \sum_{j<k} \mathrm{Cov}\left(\Delta_j, \Delta_k\right) \tag{26}$$

If prices followed a martingale along trading time, the covariance terms would be zero. The continuous-trading indices are averages over overlapping trading windows, however. Averaging induces positive correlation of first differences (Working 1960). The covariance share is therefore descriptive.

Three further metrics apply to both dimensions and test robustness to the choice of measure: the normalized mean absolute deviation, the normalized interquartile range and the normalized 99th percentile of the absolute deviation, which describes the tails. They are not trimmed. A third group of measures applies to one axis only: the root mean square of hour-to-hour changes and the mean daily range of day-ahead prices for variability, the range estimator of Parkinson (1980), based on the highest and lowest trade price of each hourly product, for volatility. Robustness checks also use the untrimmed standard deviation and a 0.5/99.5 percentile threshold. Log returns are not used, because prices are frequently zero or negative.

From the transactions, I take the last trade at least 3 h and at least 1 h before delivery and the last trade overall as point prices and compute their spread to the day-ahead price. Point prices avoid the averaging of the continuous-trading indices. For the price path within the trading session, I compute the realized variance on a 5-minute grid with the last price per interval, starting at the day-ahead price (Andersen and Bollerslev 1998), a two-scale realized variance that corrects for microstructure noise (Zhang et al. 2005), the bipower variation, which separates jumps (Barndorff-Nielsen and Shephard 2004), and the Parkinson (1980) range estimator. The annual value is the square root of the trimmed mean across products, divided by the annual mean day-ahead price. Quarter-hourly products are measured against the opening auction.

All measures are divided by the annual mean day-ahead price, on both axes. Absolute measures in €/MWh move with inflation and with general swings of energy prices, such as the gas price crisis of 2021–2022, which would dominate any comparison over 25 years. In the model, fuel and carbon prices scale variability and volatility proportionally through the slope of the merit order (Section 4.3); the ratio removes this common factor and makes measures comparable across decades and price levels. The mean price itself depends on wind and solar power: by lowering it, renewables raise relative measures. Robustness checks therefore divide by the daily mean price (days with a mean below 10 €/MWh excluded) and by the short-run marginal cost of a combined-cycle gas turbine (efficiency 50 %, 0.202 t CO2 per MWh of gas), which does not depend on renewables.

Hours are the main resolution for both concepts. Until September 2025, the day-ahead auction traded hourly products; quarter-hourly spreads against the hourly day-ahead price would therefore contain the within-hour shape, which market parties trade only in the intraday market (Kremer et al. 2021). Quarter-hours are analyzed separately, with the opening auction as the quarter-hourly base before October 2025 (Section 6.7). The same holds for the imbalance price: from October 2025, when the day-ahead auction trades quarter-hours, quarter-hourly imbalance-DA spreads serve as a sensitivity.

# 6. Results

This section first presents the normalized standard deviation of prices across both time dimensions, i.e., variability and volatility. Then I discuss drivers and present robustness checks.

## 6.1. Long-term trends in variability and volatility

Figure 5 shows the main measure on both axes, the normalized standard deviation. Variability of day-ahead prices was 24–40 % of the annual mean price in 2010–2019 and has been 50–51 % in 2024–2026, at the upper end of the range of the 2000s (34–56 %). Volatility of the imbalance price, measured by the imbalance-DA spread, fell from 76–147 % in 2010–2015 to 42–53 % in 2025–2026. The expansion of wind and solar power coincides with higher variability, not with higher volatility.

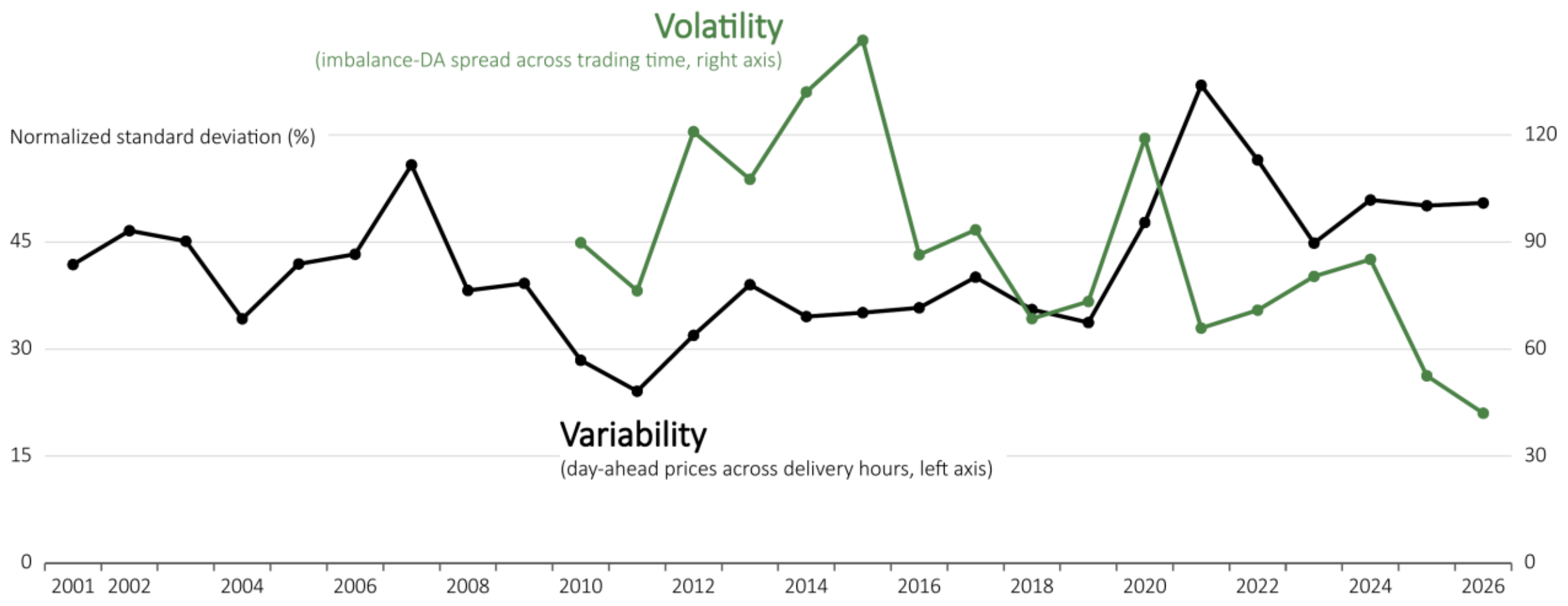

Figure 5. Variability is higher than in the 2010s; volatility against the imbalance price has fallen since 2015

## 6.2. Variability and its composition

Variability of day-ahead prices declined from the 2000s (average 43 %) to a low in the 2010s (34 %), peaked in 2021, when gas prices changed sharply from month to month, and has been at the level of the 2000s since 2024 (51 %; Figure 5). In the past three years, variability was only marginally higher than in the first three years, despite an increase of wind and solar in total generation by more than 40 percentage points.

The variance decomposition (Figure 6) shows how the composition of variability has changed. The within-day component accounted for more than half of the variance of hourly prices until 2011 and for about half in 2012–2020. In 2021 and 2022, when gas prices rose sharply and changed from month to month, variation between months dominated. Overall, the composition during the last three years is remarkably similar to the first three years.

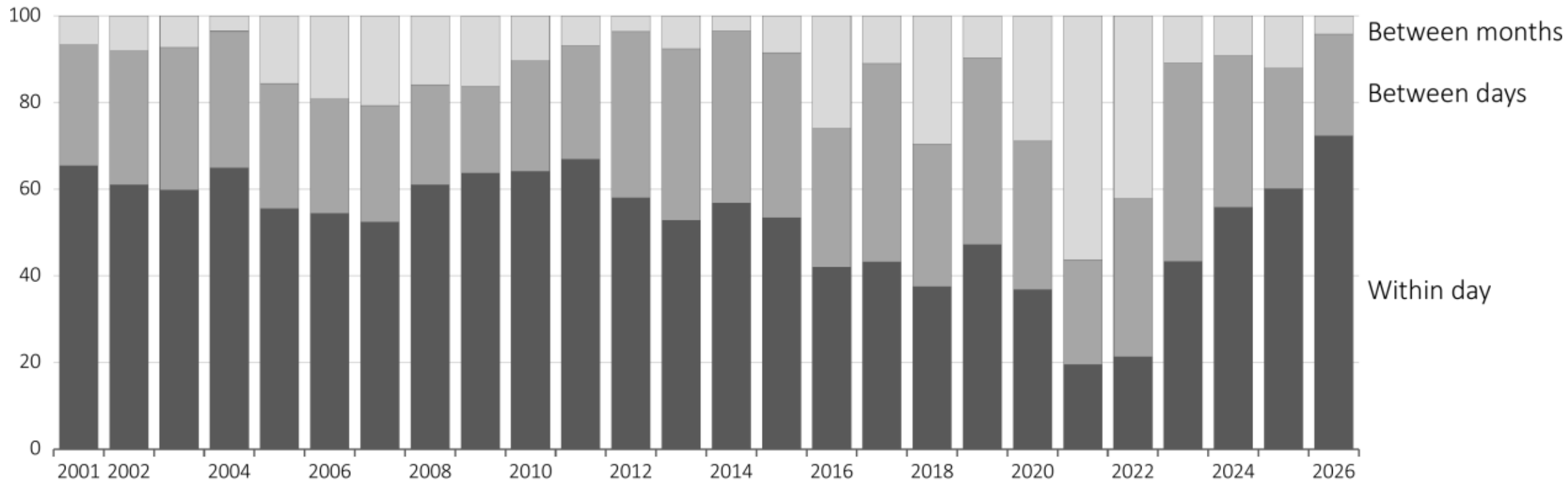

Figure 6. Decomposition of variability

Figure 7 shows alternative measures of variability, each indexed to its 2001 value. The mean absolute deviation, the interquartile range and the 99th percentile of the absolute deviation from the annual mean as well as the root mean square of hour-to-hour changes and the mean daily range fell after the 2000s and have risen since 2020. Broadly speaking, they confirm the U-shaped long-term pattern and the finding that variability today is, at best, slightly higher than 25 years ago.

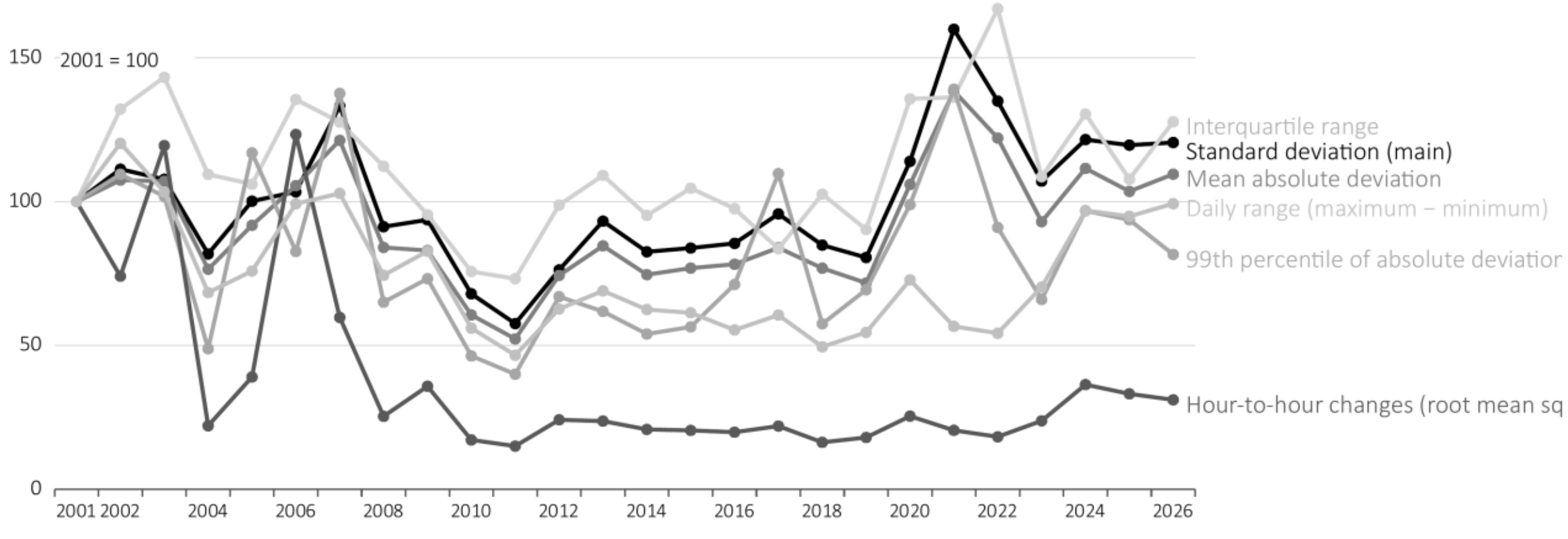

Figure 7. Alternative metrics of variability. All measures show a U-shaped long-term trend

## 6.3. Volatility: imbalance and intraday prices

Figure 8 shows the volatility of German electricity prices. None of the intraday series has a trend. Volatility increases along trading time: the opening auction deviates little from the day-ahead price, the full session, ID3 and ID1 progressively more. The highest values occurred in 2020, a year of low prices, and in 2024; the lowest in 2021 and 2022, when prices were highest. The series start when the respective trading stage became liquid enough to compute an index.

In absolute terms, volatility rose with the price level after 2021, as the model predicts for a common scaling factor (Section 4.3).

The imbalance price deviates much more from the day-ahead price (Figure 8, right axis). Volatility of the imbalance-DA spread fell from 76–147 % in 2010–2015 to 42–53 % in 2025–2026. The imbalance price is the final price of each delivery period, but it is set by a formula. Breaks coincide with changes of that formula in December 2012, May 2016, 2020 and June 2022; the series describes the settlement regime as much as price formation.

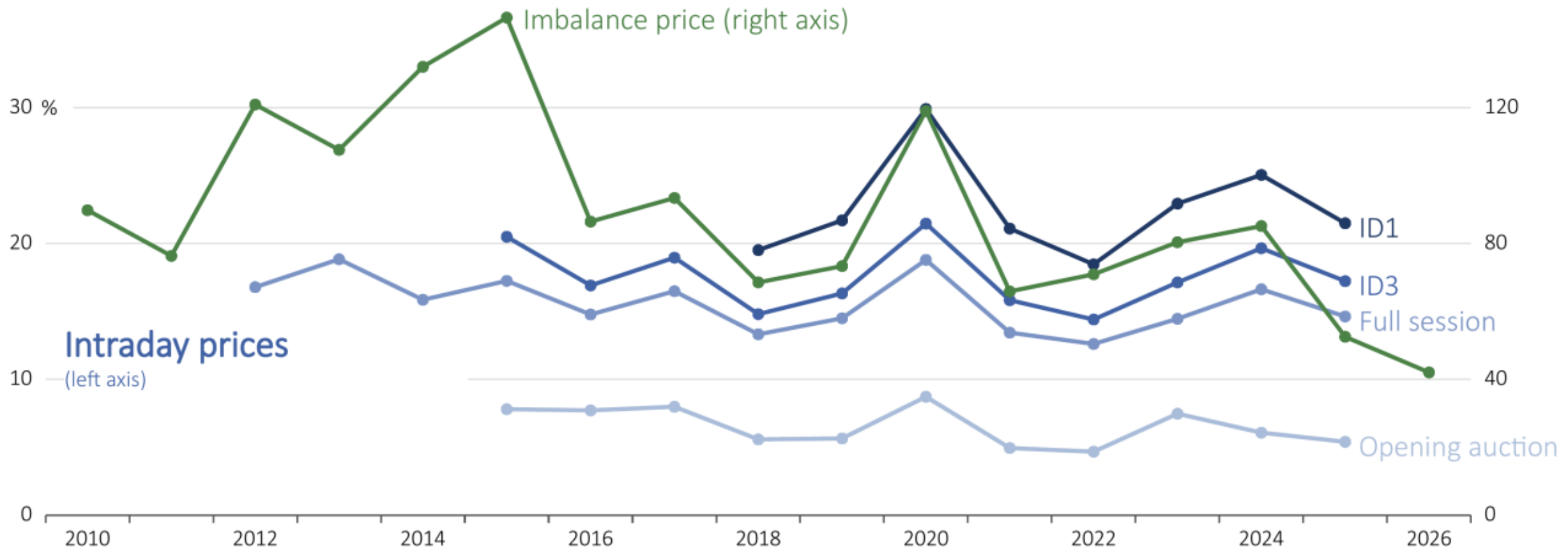

Figure 8. Intraday volatility shows no trend; imbalance price volatility has fallen since 2015

Figure 9 shows alternative measures of volatility, each indexed to its first year (2015 for the ID3-DA spread, 2010 for the imbalance-DA spread). All measures of intraday volatility except the range estimator show no trend. For the imbalance-DA spread, all four measures of dispersion have declined since 2015. The range estimator widens with the number of trades and with single outlier trades and therefore reflects the growth of liquidity as much as volatility.

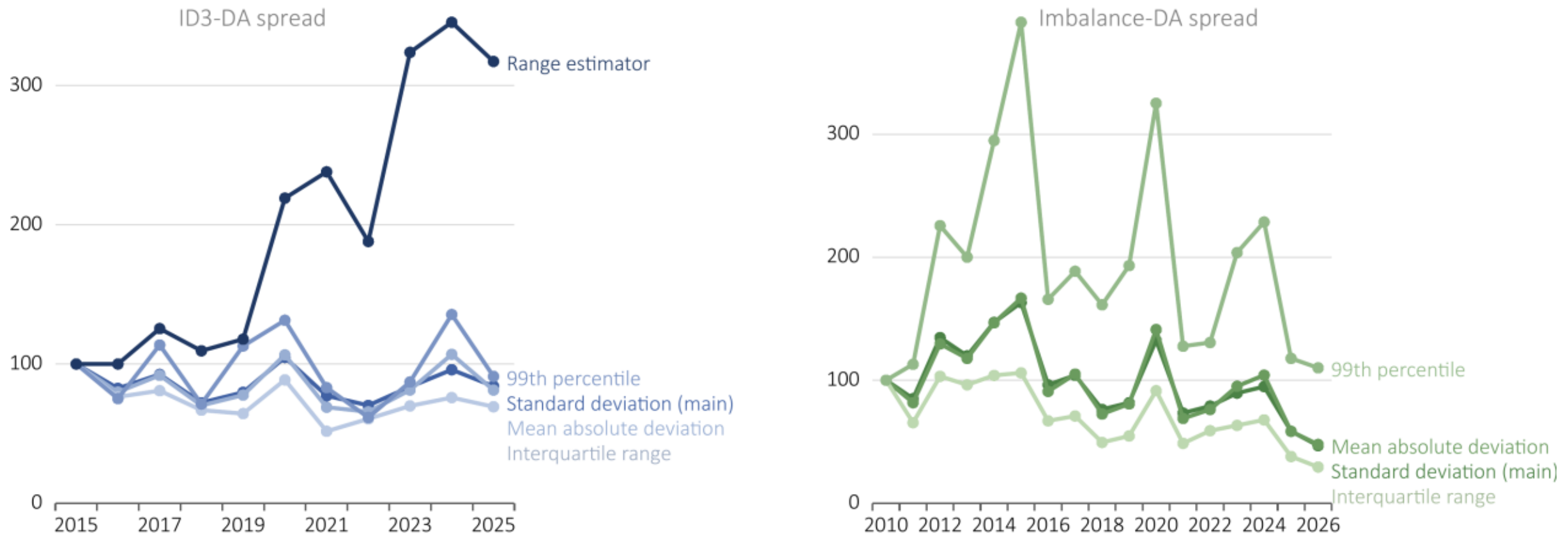

Figure 9. All measures except the range estimator show no trend in intraday volatility

Volatility builds up along trading time. Of the variance of the ID1-DA spread in 2018–2025, the step from the day-ahead auction to the opening auction accounts for 6–12 %, the step from the opening auction to the full continuous session for 36–49 %, the step to ID3 for 2–4 % and the last trading hour for 15–25 %. The remaining 22–33 % are covariance terms: consecutive increments are positively correlated. Because the continuous-trading indices average over overlapping windows, this correlation is partly mechanical (Section 5.2). This decomposition uses the common sample of hours in which all stages are available. The intraday auctions introduced in June 2024 are snapshots rather than averages. Measured against them, volatility rises monotonically along trading time, from 6–7 % at IDA1 to 9 % at IDA2, 13–15 % at IDA3 and 24–32 % at ID1 (delivery hours 12:00–24:00, which IDA3 covers, 14 June 2024 to December 2025). Information keeps arriving until shortly before delivery.

Point prices from all transactions show the same picture as the indices. The trimmed standard deviation of the spread between the last trade and the day-ahead price was 31 % of the annual mean price in 2015, 30 % in 2019 and 30 % in 2025; it was higher only in 2020, when the mean price was low. The spreads of the last trade at least 3 h and at least 1 h before delivery show no trend either. Within the trading session, the realized and the two-scale realized volatility have declined since 2015; the range estimator has risen slightly (Figure 10). The spread between the imbalance price and the last trade fell from 142 % of the annual mean price in 2015 to 43 % in 2025.

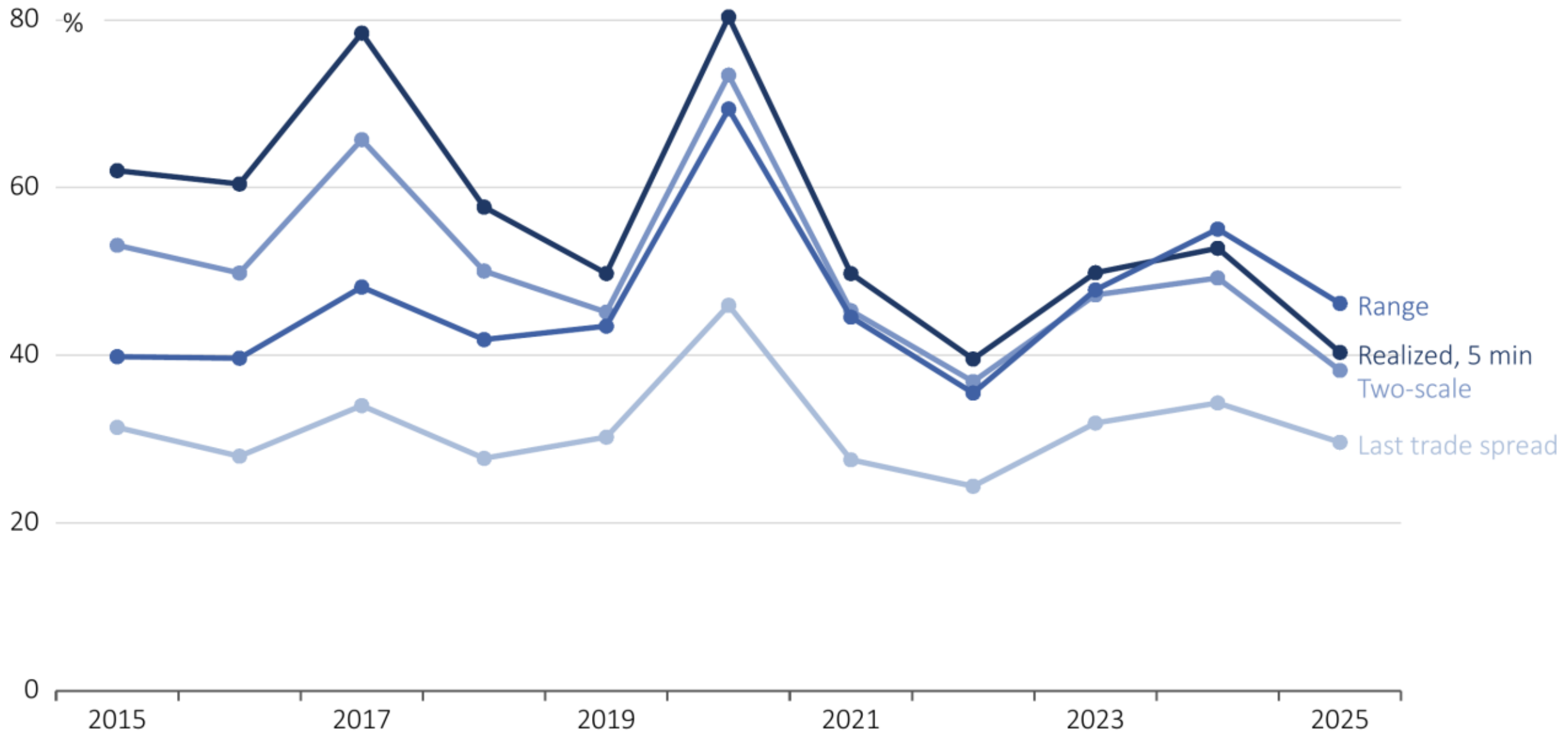

Figure 10. Transaction data: path measures within the trading session indicate declining volatility

Consecutive price changes between the day-ahead auction, the last trade at least 3 h and at least 1 h before delivery and the last trade are almost uncorrelated, with correlations between −0.12 and 0.02. On the 5-minute grid, however, the realized variance exceeds the squared spread between the last trade and the day-ahead price by a factor of 1.5–4.7 (Figure A3). Prices revert on short horizons, which indicates microstructure noise. The ratio declined as the number of trades per hourly product rose from about 260 in 2015 to about 7,000 in 2025. The share of jumps in the realized variance, measured by bipower variation, fell from 45 % to 21 %.

## 6.4. Decomposition and co-movement

Over all hours of a year, I decompose the variance of the imbalance price and of the ID3 index according to (25), divided by the squared annual mean day-ahead price. Day-ahead variability was 6–15 % in 2010–2019, 41 % in 2021 and 20–25 % in 2023–2026. The variance of the imbalance-DA spread was much larger: 46–211 % until 2020 and 42–70 % in 2021–2024. In 2025 and 2026, it fell to 26 % and 17 %, about the level of day-ahead variability. The variance of the ID3-DA spread was 2–4 %, an order of magnitude below day-ahead variability in every year. The covariance term is negative in every year for both prices; the correlation between day-ahead price and spread is −0.01 to −0.15 for the imbalance price and −0.05 to −0.20 for ID3. The spreads partly offset the day-ahead price shape. In 2024 and 2025, the variance of the ID3 price was slightly lower than day-ahead variability.

The model predicts a zero covariance term if prices carry no risk premium (Section 4.3). Three explanations for the negative sign are consistent with the data: a state-dependent risk premium, additional flexibility in the intraday market that responds to the day-ahead shape and the construction of ID3 as an average over three trading hours. The present data do not allow separating them. For the imbalance price, the settlement formula is a further candidate.

Monthly variability and volatility are strongly correlated (Figure 11). Each dot is one month of 2016–2025; both measures are root mean square deviations from the daily mean, pooled over the hours of a month. In the model, the co-movement reflects the common slope of the merit order, not a causal link between the two measures.

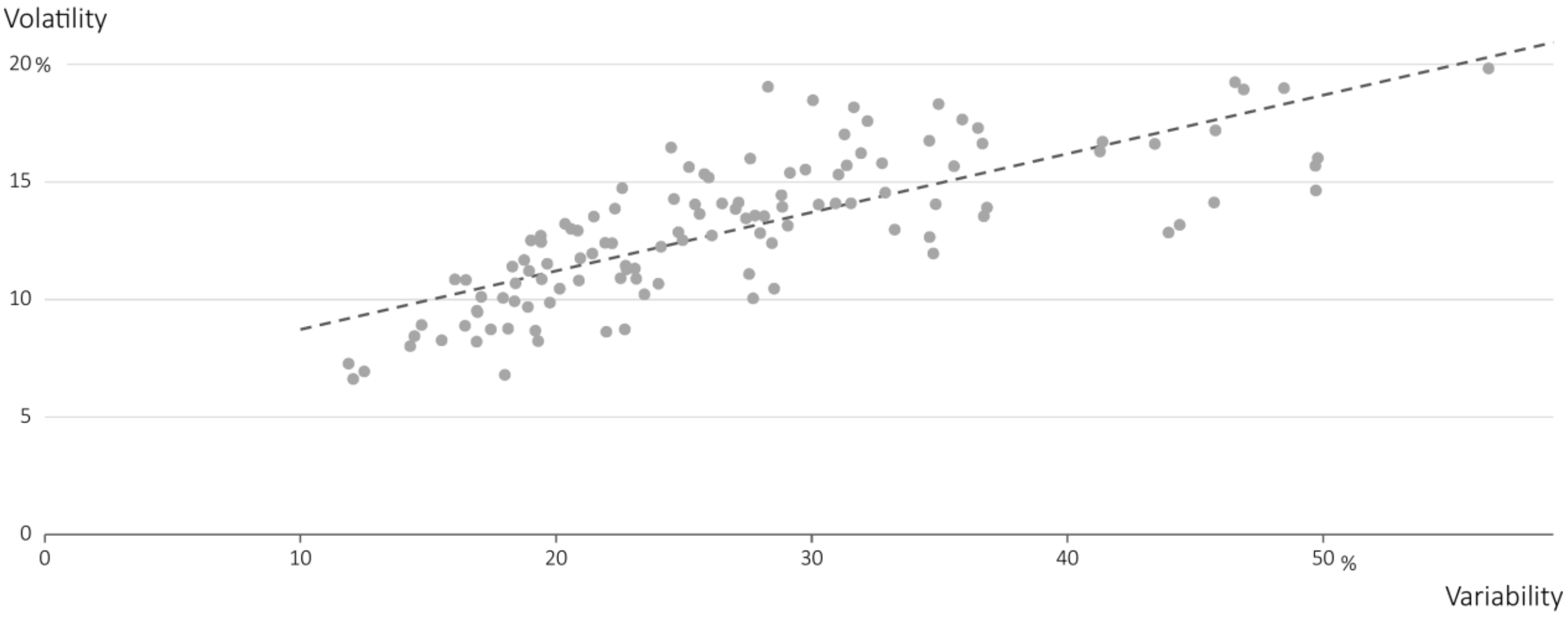

Figure 11. Monthly variability and volatility move together

## 6.5. The role of wind and solar

Figure 12 shows within-day variability of day-ahead prices by daily mean solar generation (left) and wind generation (right), one line per year; days are grouped in bins of 2 GW of solar and 5 GW of wind generation, bins with fewer than ten days are omitted. Until 2020, variability was highest on days with little solar generation: a small solar fleet flattens the load-driven price shape. Since 2024, variability rises steeply with solar generation. This matches the turning point in the model (8): solar power first flattens the price shape and then reverses it. Wind power has the opposite effect in recent years: until 2020, variability rose slightly with wind generation; since 2024, the windiest days have the lowest variability.

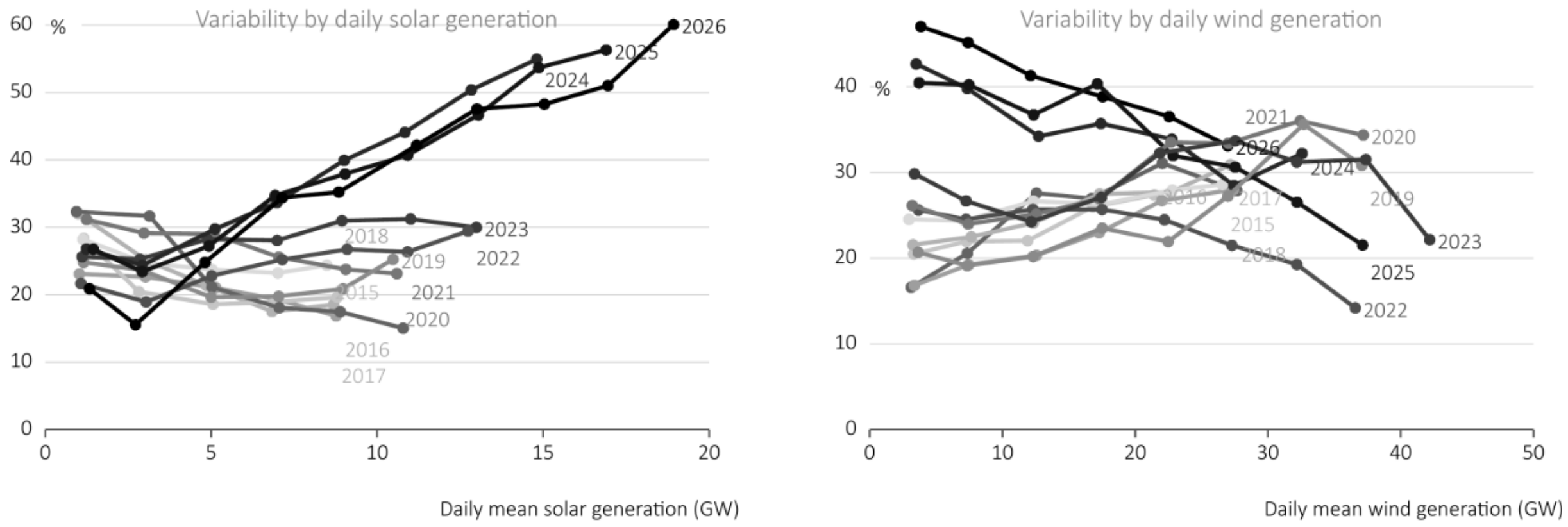

Figure 12. Since 2024, sunny days have the highest and windy days the lowest variability

Measured by the imbalance-DA spread, volatility rose with daily wind generation in most years until 2020 and in 2023; in 2021, 2022 and 2024–2026, it hardly depends on wind generation (Figure 13). Solar generation shows no consistent pattern. Differences between years are larger than those within a year. Actual generation describes the predictable component of wind and solar power, not the forecast errors that drive volatility in the model; forecast errors of market parties are not publicly available.

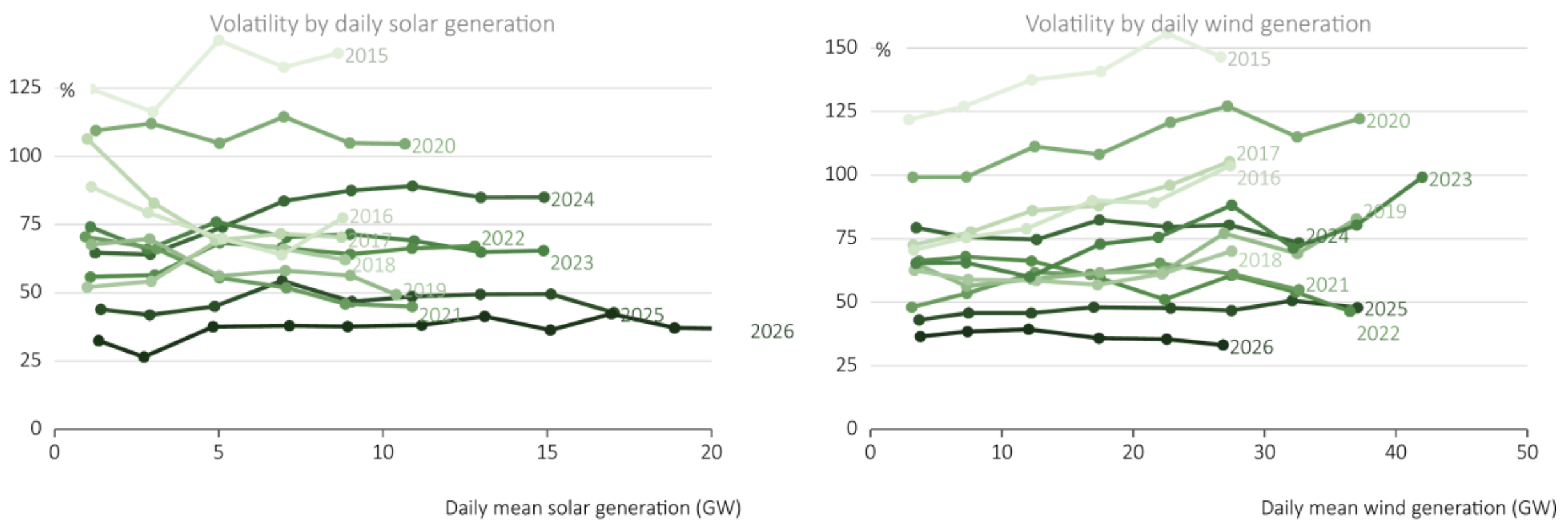

Figure 13. Volatility no longer rises with wind generation

## 6.6. Three drivers of volatility

The development of German volatility over time reflects three drivers, which this paper describes but does not separate. First, fundamentals: the capacity of wind and solar power and the size of their forecast errors, the level and shape of load and the flexibility of the system, including storage, which dampens volatility (Section 4.4). Second, liquidity: continuous intraday trading volume in Germany grew more than fivefold between 2012 and 2025. Continuous trading has been coupled across Europe since 2018. More liquidity can itself lower volatility, because a given forecast error meets more counterparties and more flexible capacity; index values of the early years rest on few trades. Third, the regulation of the imbalance price: reforms of the settlement formula change how strongly the imbalance price reacts to a given forecast error and thereby what intraday prices anticipate. The breaks in the imbalance price series (Figure 8) are consistent with this channel.

Separating the three drivers requires a regression framework with fundamentals, measures of liquidity and the dates of regulatory reforms. I leave this to future research.

## 6.7. Quarter-hours

Quarter-hourly prices vary considerably around their hourly mean, in the opening auction and in the ID3 index (Figure A2, left); the measure is the trimmed root mean square deviation of quarter-hourly prices from their hourly mean. Since October 2025, the day-ahead auction itself trades quarter-hours; its intra-hour variability is lower than that of the intraday prices. The series is too short for conclusions on the effect of the change.

Measured against the same base, the opening auction, the volatility of quarter-hours exceeds the volatility of hourly means in every year (Figure A2, right). Averaging to hours understates the volatility of individual quarter-hours, but the development over time is the same. The same holds for the imbalance price: from October 2025 to October 2026, the normalized standard deviation of the imbalance-DA spread is 47 % for quarter-hours and 41 % for hourly means.

## 6.8. Robustness: trimming and normalization

A trimming threshold of 0.5/99.5 raises levels slightly without changing the pattern (Figure 14). Without trimming, volatility shows peaks in 2015, 2017, 2020 and 2024 that stem from a few extreme days; variability in the early 2000s is dominated by individual spikes.

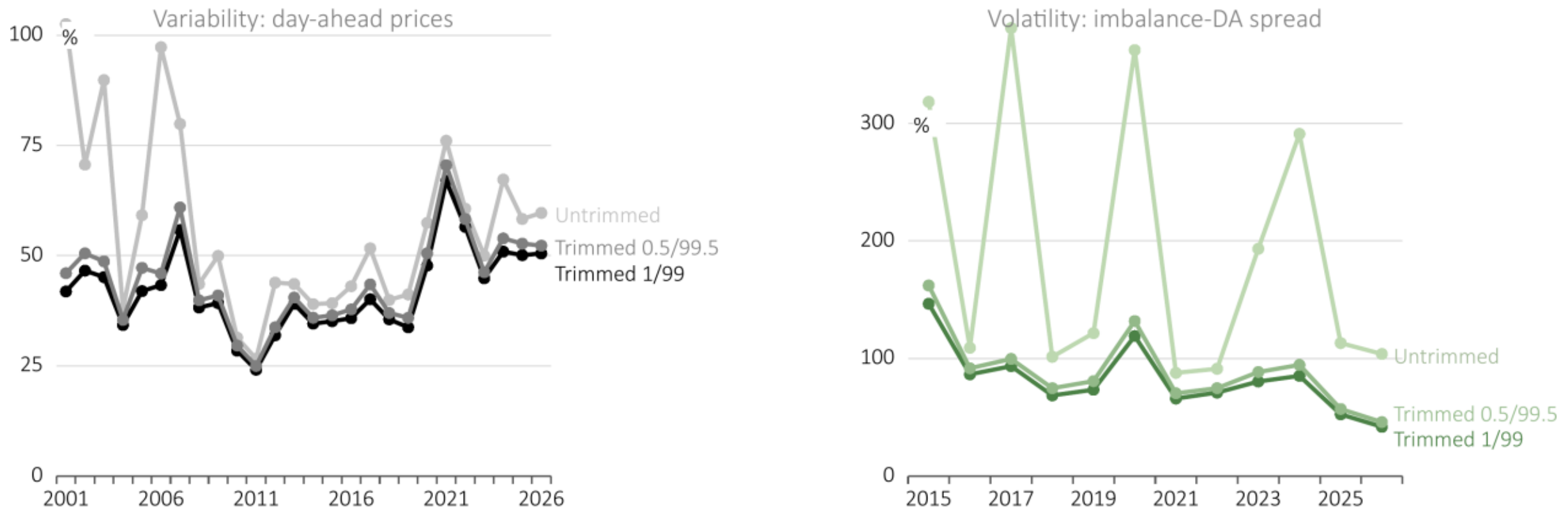

Figure 14. Results are robust to trimming of data

Normalized by the marginal cost of a gas-fired plant instead of the annual mean price, variability shows the same pattern until 2023; carbon prices for 2024–2026 are not yet included (Figure A1). For the imbalance-DA spread, all three normalizations show the decline since 2015; divided by the daily mean price, volatility is higher in every year, because days with low mean prices and large spreads receive more weight.

## 6.9. Price variability across Europe

Figure 15 shows variability for eight European bidding zones since 2015, measured as the trimmed standard deviation of hourly day-ahead prices around the annual mean, divided by the annual mean day-ahead price. Variability has risen in almost all bidding zones. In 21 of the 22 zones with data for 2015–2019, it was higher in 2024–2026 than in 2015–2019; the median across zones rose from 30 % to 52 %. The exception is Northern Italy. The increase was largest in Finland, Spain, Portugal and Sweden (SE3), by 50–70 percentage points. In Germany, variability rose by 14 percentage points.

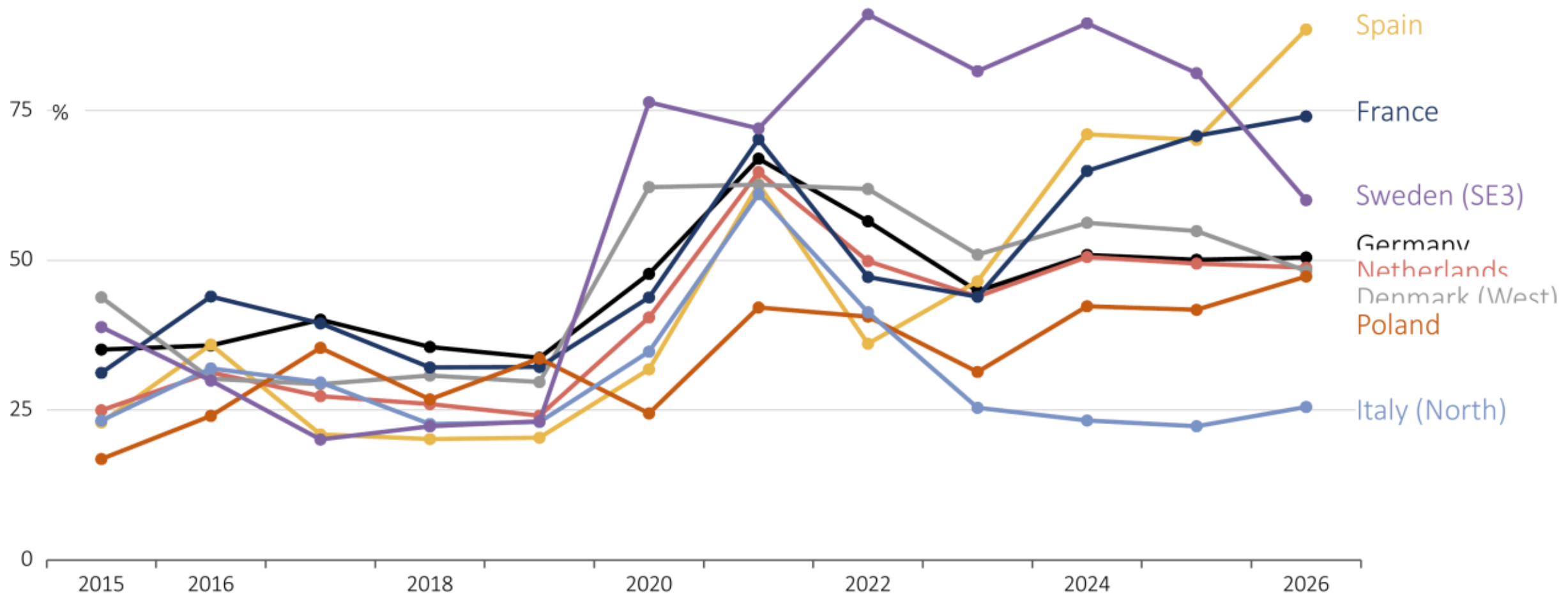

Figure 15. Variability has risen in almost all European bidding zones

Monthly variability rises with the monthly share of solar and wind energy (Figure 16).

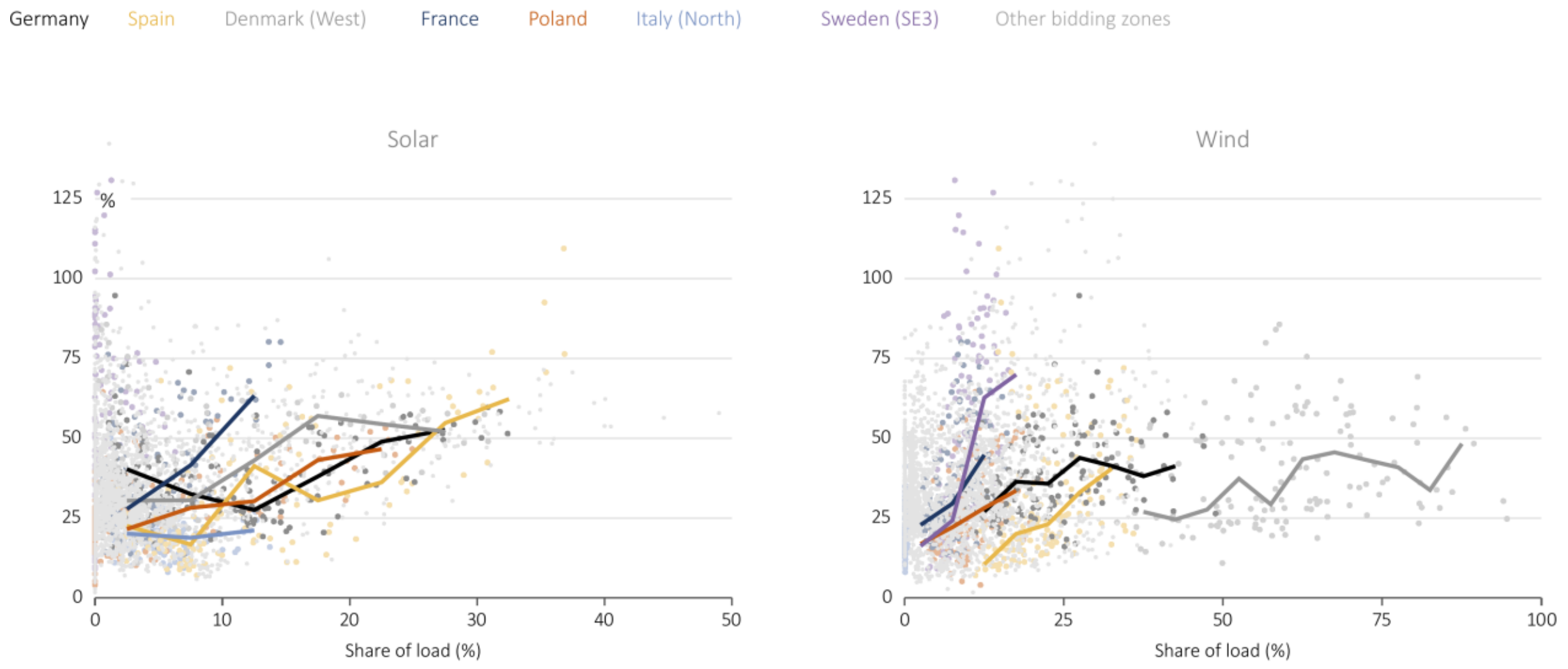

Figure 16. Month by month: variability vs. renewables

# 7. Conclusions

The common statement that wind and solar power increase electricity price volatility conflates two phenomena: variability, price differences across delivery periods, and volatility, price changes of the same delivery period across trading time. The former is the result of non-storability, the latter of uncertainty. Most studies on renewables and price "volatility" measure variability.

My preferred measure of variability is the trimmed, normalized standard deviation. By this measure, the variability of day-ahead prices in Germany *fell* from the 2000s to the 2010s, despite a massive increase in wind and solar power. It has increased since then, but only to marginally above the 2001–2003 level. Alternative metrics show a similar U-shaped pattern over time. The decomposition into within-day, between-day and between-month variability yields similar shares as 25 years ago.

This paper measures volatility as the trimmed, normalized standard deviation of the spread between day-ahead and imbalance or intraday prices. The data comprise imbalance prices, intraday price indices and transaction data. Volatility in the German intraday market has shown no trend since 2012; volatility of the imbalance price has markedly declined since 2015, with 2026 showing the lowest level by far.

The analysis is descriptive. Time trends do not separate the effect of wind and solar power from those of fuel prices, market integration, regulation and new flexibility. I leave it for future research to identify and disentangle the drivers of variability and volatility.

For risk management, the distinction matters because the two risks call for different instruments. Variability is a source of shape (profile) risk, which is hedged with products for specific delivery periods and creates the spreads that storage arbitrages. Volatility is a risk of price changes before delivery, giving rise to balancing risk. German price data do not show a significant increase of either one as the wind and solar market share increased from close to zero to nearly 50 %.

# Appendix

Figures A1–A3 show supplementary results for Germany: robustness to the normalization (Figure A1), quarter-hourly variability and volatility (Figure A2) and liquidity and microstructure noise in continuous trading (Figure A3).

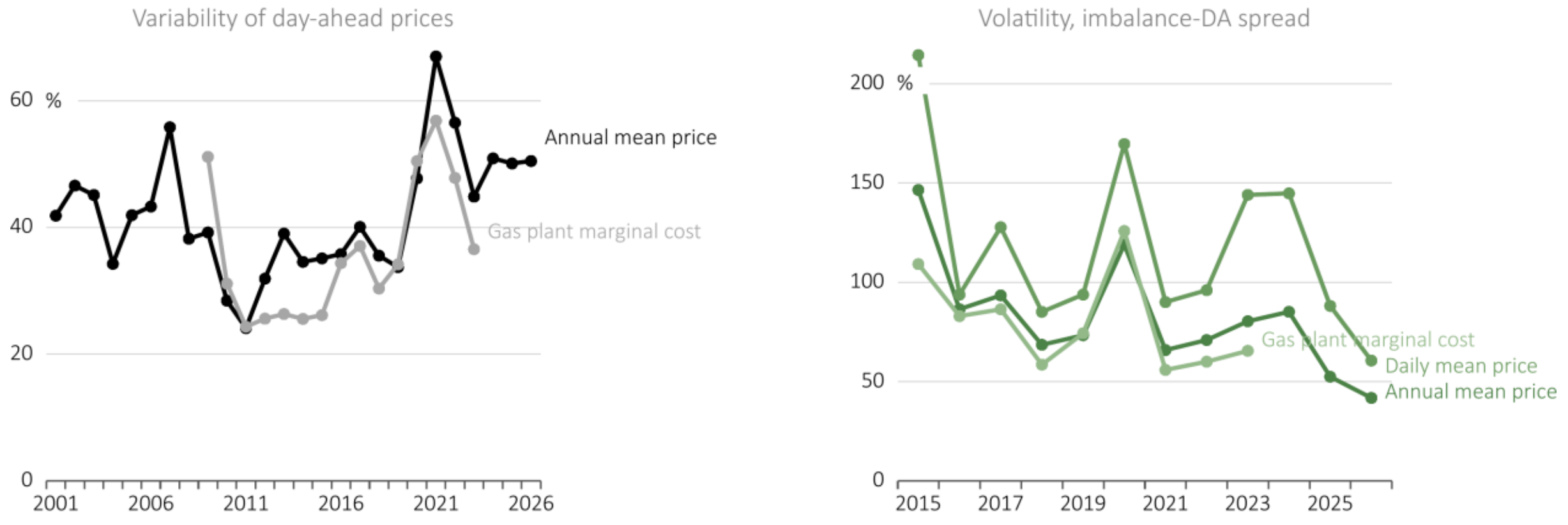

Figure A1. The pattern does not depend on the normalization

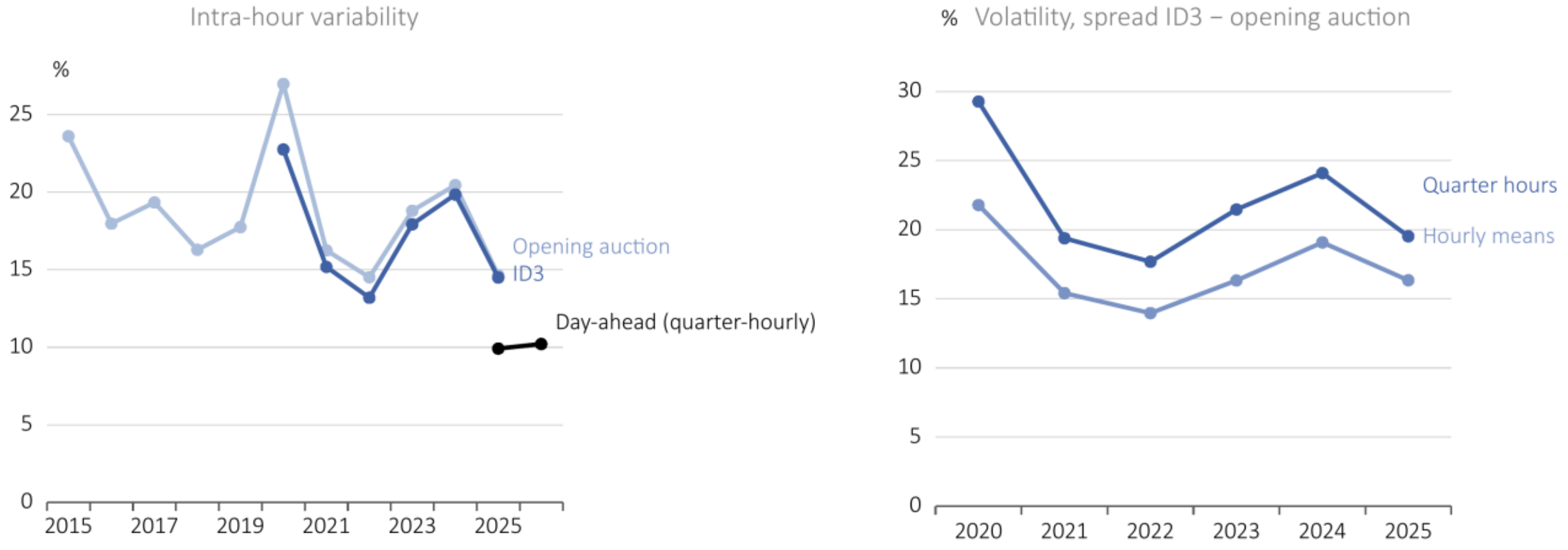

Figure A2. Averaging to hours understates the level, not the development, of quarter-hourly volatility

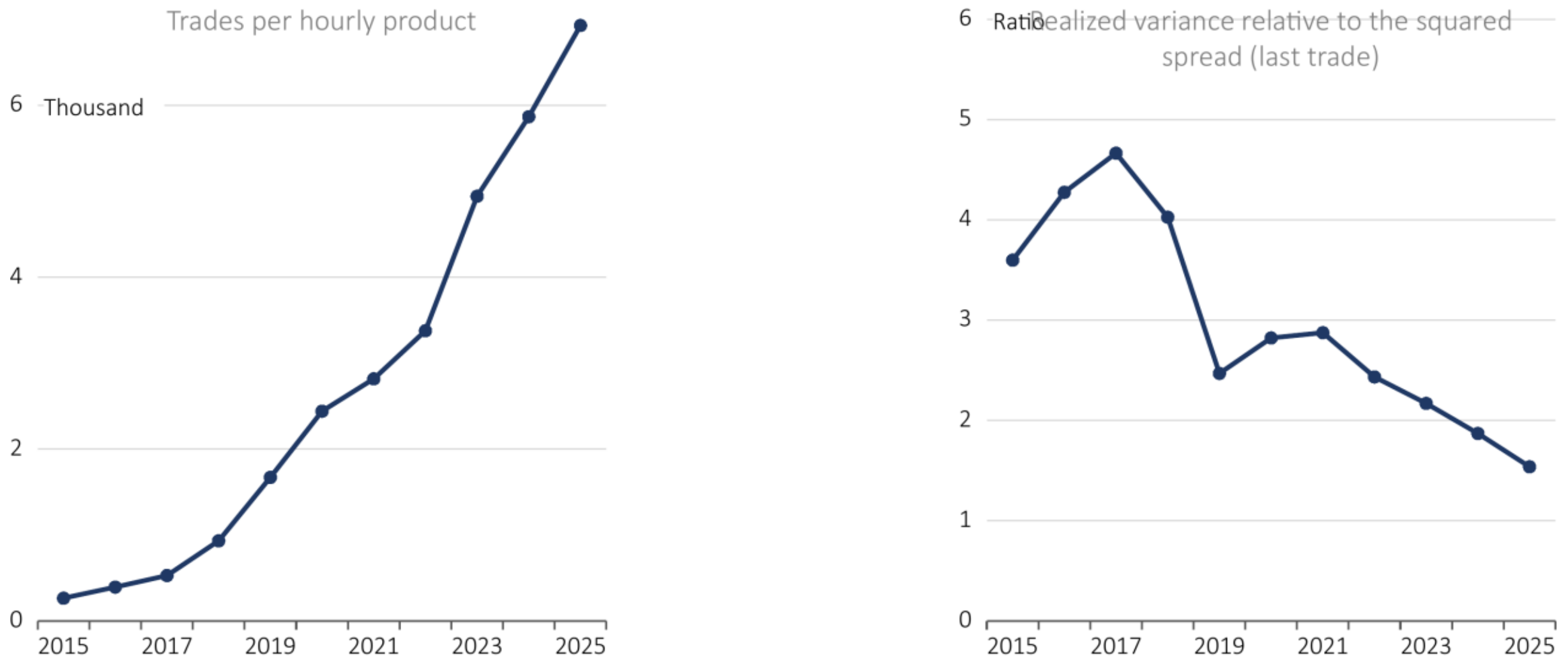

Figure A3. More trades, less noise in the price path

# Statements

## Acknowledgements

I thank Ingmar Schlecht for the countless discussions on price volatility and variability we have had over the past ten years. I have learned more from him than from anyone else.

## Competing interest

The author is managing director of Neon Neue Energieökonomik, a consultancy advising clients in the electricity sector. This paper did not receive any funding and was not commissioned by any client; no client has seen its results prior to publication.

## Funding

This research did not receive any specific grant from funding agencies in the public, commercial, or not-for-profit sectors.

## Data availability

Day-ahead, imbalance and generation data are publicly available from the sources listed in Table 5. EPEX SPOT intraday data are proprietary and were used under license. Code to reproduce all measures and figures is available from the author on request.

## CRediT contribution statement

Lion Hirth: Conceptualization, Methodology, Software, Formal analysis, Investigation, Writing – original draft, Writing – review & editing, Visualization.

## Declaration of generative AI in the writing process

During the preparation of this work the author used Claude (Anthropic) in order to draft and edit text, write analysis code, and produce tables and figures. After using this tool, the author reviewed and edited the content as needed and takes full responsibility for the content of the published article.